\documentclass[preprint2]{aastex}

\slugcomment{Submitted to {\sl The Astrophysical Journal}}

\shorttitle{Ram pressure stripping and galaxy orbits}
\shortauthors{Vollmer et al.}

\begin{document}

\title{Ram pressure stripping and galaxy orbits:\\
The case of the Virgo cluster}

\author{B. Vollmer}
\affil{Max-Planck-Institut f\"{u}r Radioastronomie, Bonn, Germany and \\
Observatoire de Paris, Meudon, France.}
\email{bvollmer@mpifr-bonn.mpg.de}

\author{V. Cayatte, C. Balkowski}
\affil{Observatoire de Paris, DAEC, UMR 8631, CNRS et Universit\'e Paris 7,\\
F-92195 Meudon Cedex, France.}
\email{Veronique.Cayatte@obspm.fr, Chantal.Balkowski@obspm.fr}

\and

\author{W.J. Duschl\altaffilmark{1}}
\affil{Institut f\"ur Theoretische Astrophysik der Universit\"at Heidelberg,
Tiergartenstra{\ss}e 15, D-69121 Heidelberg, Germany.}
\email{wjd@ita.uni-heidelberg.de}

\altaffiltext{1}{also at: Max-Planck-Institut f\"ur Radioastronomie, 
Auf dem H\"ugel 69, D-53121 Bonn, Germany}

\begin{abstract}
We investigate the role of ram pressure stripping in the Virgo cluster
using N-body simulations. Radial orbits within the Virgo cluster's
gravitational potential
are modeled and analyzed with respect to ram pressure stripping.
The N-body model consists of 10\,000 gas cloud complexes which can have 
inelastic collisions. Ram pressure is modeled as an additional 
acceleration on the clouds located at the surface of the gas
distribution in the direction of the galaxy's motion within the
cluster. We made several simulations changing the orbital
parameters in order to recover different stripping scenarios
using realistic temporal ram pressure profiles. We investigate systematically
the influence of the inclination angle between the disk and the 
orbital plane of the galaxy on the gas dynamics. We show that ram 
pressure can lead to a temporary 
increase of the central gas surface density. In some cases a considerable part of
the total atomic gas mass (several $10^{8}$~M$_{\odot}$) 
can fall back onto the galactic disk after the stripping event.
A quantitative relation between the orbit parameters and the resulting
H{\sc i} deficiency is derived containing explicitly the inclination angle between
the disk and the orbital plane.
The comparison between existing H{\sc i} observations and the results
of our simulations shows that the H{\sc i} deficiency depends strongly
on galaxy orbits. It is concluded that the scenario where ram pressure stripping 
is responsible for the observed H{\sc i} deficiency is consistent
with all H{\sc i} 21cm observations in the Virgo cluster.
\end{abstract}


\section{Introduction}

Since Gunn \& Gott (1972) have introduced the concept of ram pressure 
stripping, which can affect galaxies moving inside the Intracluster Medium 
(ICM) of a galaxy cluster, this mechanism has been invoked to explain 
different observational 
phenomena as the HI deficiency of spiral galaxies in clusters 
(Chamaraux et al. 1980 , Bothun et al. 1982, Giovanelli \& Haynes 1985) 
or the lower star formation activity of cluster spiral galaxies 
(e.g. Dressler et al. 1999, Poggianti et al. 1999).

Some other mechanisms have been proposed affecting spiral galaxies
in clusters, among which:
\begin{itemize}
\item 
galaxy--galaxy interactions (galaxy ``harassment''; Moore et al. 1996, 1998)
\item
galaxy--cluster gravitational interaction (Byrd \& Valtonen 1990, Valluri 1993)
\end{itemize}
All these kinds of interaction have in common that they can in principle 
remove the neutral gas from a galaxy inducing a decrease of its star 
formation activity.

The strength of these interactions depends crucially on galaxy orbits.
Radial orbits allow galaxies to go deeper into the cluster
potential where their velocity increases considerably and where the
galaxy density and the density of the ICM is substantially higher. 
Dressler (1986) analyzed Giovanelli \& Haynes (1985) data 
showing that H{\sc i} deficient spiral galaxies are on radial orbits.
Numerical simulations (Ghigna et al. 1998) showed that
galaxy halos evolving within the cluster settle on isotropic orbits with
a median ratio of pericentric to apocentric radii of 1:6.

The recent analysis of a spectroscopic catalog of galaxies in ten distant 
clusters (Dressler et al. 1999, Poggianti et al. 1999) has shown that the 
galaxy populations of these clusters are characterized by the presence of a 
large number of post starburst galaxies. Poggianti et al. (1999) concluded that 
the most evident effect due to the cluster environment is the quenching of 
star formation rather than its enhancement. They found two different 
galaxy evolution timescales in clusters. 
(i) A rapid removal of the gas ($\sim$ 1 Gyr).
(ii) A slow transformation of morphology (several Gyr). They stated that the 
mechanism responsible for the fast gas removal without changing the galaxy's 
morphology is very probably ram pressure stripping.
Furthermore, Couch et al. (1998), using {\it HST} data, claimed that there are 
fewer disturbed galaxies than predicted by the galaxy harassment model.
They suggested that ram pressure stripping by the hot ICM truncates star 
formation.

In the local universe, the existence of a large number of post starburst 
galaxies in the Coma cluster seems to indicate a triggering of the star 
formation activity by effects which are related to the cluster environment 
(Bothun \& Dressler 1986, Caldwell et al. 1993).
Caldwell \& Rose (1997) observed an enhanced
number of post starburst galaxies within cluster environments.
On the contrary, Zabludoff et al. (1996) found that interactions with the cluster 
environment, in the form of the intracluster medium or cluster potential, 
are not essential for "E+A" formation. 
{\it HST} images of the Coma starburst and post starburst galaxies 
(Caldwell, Rose, \& Dendy 1999)
revealed that star formation takes mainly place in the
inner disk of these galaxies and is suppressed in the outer disk. 
If the starburst is related to the cluster environment,
ram pressure stripping can naturally account for these characteristics.

The best place to study the gas removal due to ram pressure
is the Virgo cluster as it is the closest cluster which can
be observed in great detail. Most of the spiral galaxies seem to have entered
the cluster only recently (within several Gyr, Tully \& Shaya 1984).
About half of them became H{\sc i} deficient (Giovanelli \& Haynes 1985).
Their H{\sc i} disk sizes are considerably reduced 
(van Gorkom \& Kotanyi 1985, Warmels 1988, Cayatte et al. 1990, 1994). 
Concerning the stellar content, their intrinsic color indices
are not significantly different from field galaxies of the same
morphological type (Gavazzi et al. 1991, Gavazzi et al. 1998).
But despite the H{\sc i} deficiency, cluster galaxies do not show a
reduced CO content (Kenney \& Young 1989, Boselli et al. 1997) neither
a reduced infrared luminosity (Bicay \& Giovanelli 1987).
This means that the neutral gas was removed without 
affecting the molecular component and without changing colors very much. 
The only interaction cited above producing 
this signature is ram pressure stripping.

Very few simulations have been done to quantify ram pressure stripping 
using Eulerian hydrodynamic (Takeda, Nulsen, \& Fabian 1984, 
Gaetz, Salpeter \& Shaviv 1987; Balsara, 
Livio, \& O'Dea 1994) or SPH codes (Tosa 1994; Abadi, Moore, \& Bower 1999).
We use a sticky particles model in which each particle 
represents a cloud complex with a given mass dependent radius. 
The viscosity of the clumpy ISM is due to inelastic collisions
between the particles. The effect of ram pressure is modeled as an 
additional acceleration on the particles located in the direction of the
galaxy's motion.

This article is devoted to the question if ram pressure stripping 
can account for the observed H{\sc i} deficiency, H{\sc i} distributions, and
velocity fields of spiral galaxies in the
Virgo cluster. Furthermore we investigate possible 
star formation mechanisms related to a ram pressure stripping event. 
We first discuss radial galactic orbits within the cluster
gravitational potential (Section~2). The model simulating the neutral gas
content of an infalling galaxy is described in Section~3, followed
by the detailed description of the simulations (Section~4).
The evolution of the stripped gas in the intracluster medium is
investigated in Section~5.
In Section~6 we give the results of these simulations and discuss them
in Section~7. We confront observations of various quantities of
spiral galaxies with the knowledge acquired with the help of the simulations
(Section~8) and give possible explanations for their dynamical
and physical state (Section~9 and 10). The summary and conclusions
are given in Section~11.

We adopt a velocity of 1150 km\,s$^{\rm -1}$ for the Virgo 
cluster (Huchra 1988) and a distance of 17~Mpc.

\section{\label{sec:orbits} Galaxy orbits}

The hot intracluster medium (ICM) of density $\rho_{\rm ICM}$ 
exerts a ram pressure: $p_{\rm ram}=
\rho_{\rm ICM}\,v_{\rm galaxy}^{2}$ on a galaxy, which moves 
with a velocity $v_{\rm galaxy}$ through it.
With decreasing distance to the cluster center the galaxy's velocity,
the ICM density, and thus ram pressure increase.
Therefore, the efficiency of ram pressure stripping depends strongly
on the shape of the galaxy orbit within the cluster.
Radial orbits lead a galaxy deeper into the cluster core where ram pressure
stripping is the most important. We investigate in this section the
influence of the shape of radial orbits on the efficiency of
ram pressure stripping. 

\begin{figure}
	\resizebox{\hsize}{!}{\includegraphics{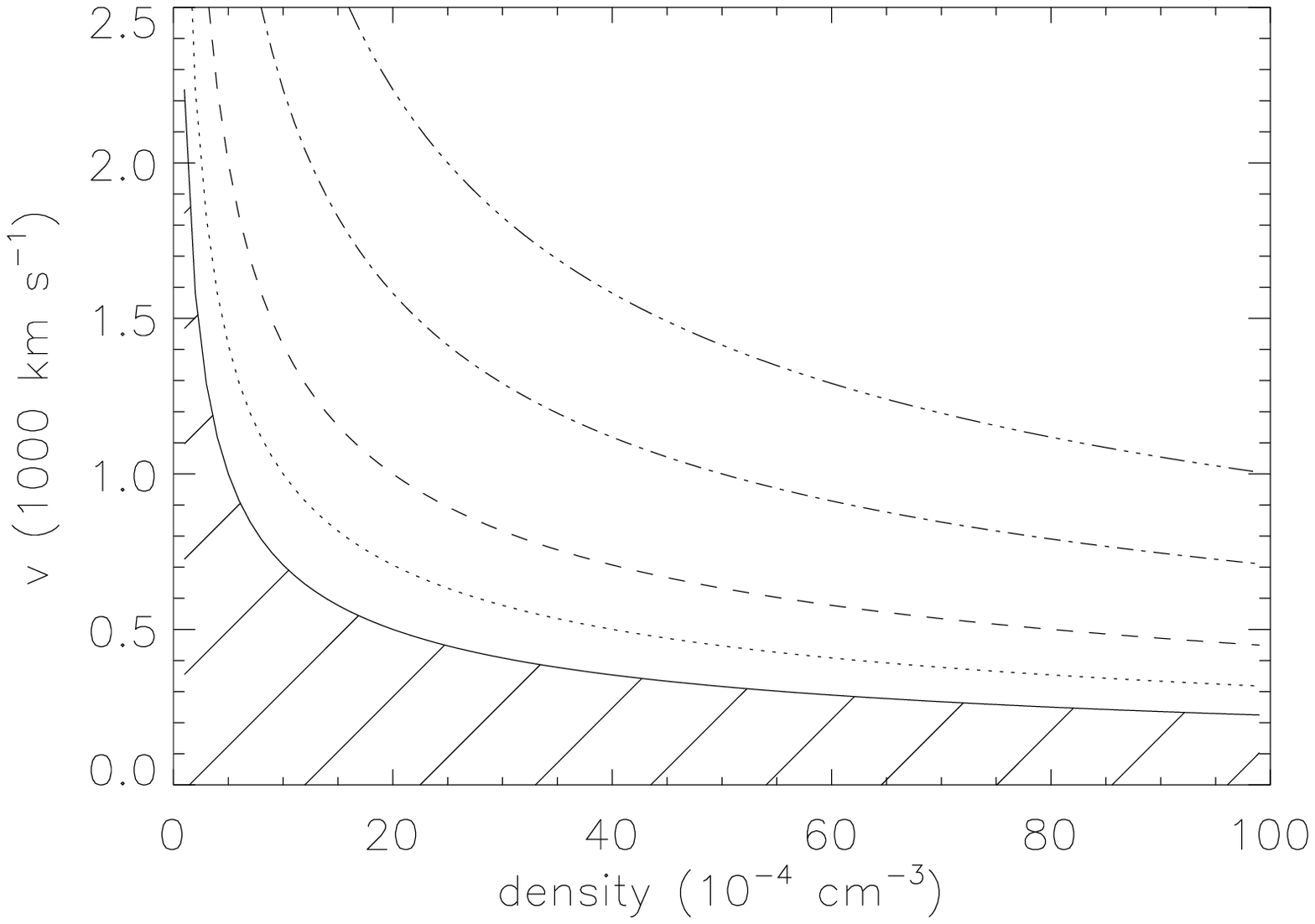}}
	\figcaption[figure1.ps]{The galaxy velocity as a function of the ICM density for a given
ram pressure $p_{\rm ram}=\rho_{\rm ICM} v_{\rm galaxy}^{2}$. Solid line:
the minimum ram pressure to affect the atomic gas disk 
$p_{\rm ram}=0.5\times 10^{3}$~cm$^{-3}$(km\,s$^{-1}$)$^{2}$. The area below the solid 
line corresponds to the part of the parameter space where ram pressure
is not effective. Dotted line: 
$p_{\rm ram}=10^{3}$~cm$^{-3}$(km\,s$^{-1}$)$^{2}$. Dashed line: 
$p_{\rm ram}=2\times 10^{3}$~cm$^{-3}$(km\,s$^{-1}$)$^{2}$. Dot-dashed line:
$p_{\rm ram}=5\times 10^{3}$~cm$^{-3}$(km\,s$^{-1}$)$^{2}$. Dot-dot-dashed line:
$p_{\rm ram}=10^{4}$~cm$^{-3}$(km\,s$^{-1}$)$^{2}$.
\label{fig:parameterspace}}
\end{figure}

The gas in the galactic disk is removed if the ram pressure of the ISM
is greater than the restoring gravitational force per unit area provided 
by the galaxy's disk (Gunn \& Gott 1972). 
The stripping radius $R_{\rm str}$ can be estimated using the equation established by
Gunn \& Gott (1972) with $\Sigma_{*}=(2\pi G)^{-1}\,v_{\rm rot}^{2} R^{-1}$
(Binney \& Tremaine 1987):
\begin{equation}
\Sigma_{\rm gas}\,v_{\rm rot}^{2} R^{-1}=\rho_{\rm ICM} v_{\rm galaxy}^{2}\ ,
\label{eq:gunngott}
\end{equation}
where $\Sigma_{*/{\rm gas}}$ is the stellar/gas
surface density, $\rho_{\rm ICM}$ is the ICM density, $v_{\rm rot}$ is the 
rotation velocity, and $v_{\rm galaxy}$ is the galaxy's 
velocity within the cluster. We assume 
$\Sigma_{\rm gas}={\rm const}=10^{21}$~cm$^{-2}$ and $v_{\rm rot}=150$~km\,s$^{-1}$.
The minimum ram pressure
$p_{\rm ram}^{\rm min}$ to affect the atomic gas disk can be determined by setting 
$R=R_{\rm HI}=15$~kpc. For our model galaxy this leads to 
$p_{\rm ram}^{\rm min}\sim 500$~cm$^{-3}$\,(km\,s$^{-1}$)$^{2}$.
Fig.~\ref{fig:parameterspace} shows the galaxy velocity as a function of the ICM density
for constant values of the ram pressure $p_{\rm ram}=\rho_{\rm ICM}\,v_{\rm galaxy}^{2}$.
The area below the solid line corresponds to the part of the parameter space where
$\rho_{\rm ICM} v_{\rm galaxy}^{2}<$500~cm$^{-3}$\,(km\,s$^{-1}$)$^{2}$, i.e.
where ram pressure is not effective. The curves correspond to a values
of $\rho_{\rm ICM} v_{\rm galaxy}^{2}$=1000, 2000, 5000, 
10000~cm$^{-3}$\,(km\,s$^{-1}$)$^{2}$
that we used in our simulations. For the simulations we varied the inclination angle $\Theta$
between the disk and the orbital plane: 
$\Theta$=0$^{\rm o}$, 20$^{\rm o}$, 45$^{\rm o}$, 90$^{\rm o}$.
With this definition of the inclination angle, $\Theta$=0$^{\rm o}$ means
edge-on stripping and $\Theta$=90$^{\rm o}$ face-on stripping.
For our model, this represents the physically most meaningful definition and
has the advantage that it leads to simple fitting functions of our numerical results.

We took a $\beta$-model (Cavaliere \& Fusco-Femiano 1976)
to describe the total mass density and the ICM density:
\begin{equation}
\rho=\rho_{0}\big(  1+ \frac{r^{2}}{r_{\rm C}^{2}}\big) ^{-\frac{3}{2}\beta}
\end{equation} \label{eq:betamodel}
where $r_{\rm C}$ is the core radius, $\beta$ is the slope parameter, and
$\rho_{0}$ is the central density. 
The values for the ICM are:
$\beta$=0.5, $r_{\rm C}$=13.4 kpc, and $\rho_{0}=4\times 10^{-2}$ cm$^{-3}$.
For the total mass profile we used 
$\beta$=1,
$r_{\rm C}$=0.32 Mpc, and $\rho_{\rm C}=3.76\times 10^{-4}$ M$_{\odot}$pc$^{-3}$.
This gives a total mass of $M_{\rm tot}=1.4\times 10^{14}$ M$_{\odot}$ at
a radius of 1 Mpc (Schindler, Bingelli, \& B\"{o}hringer 1999). 
We have calculated orbits of test particles within the gravitational
potential given by this mass distribution.

The giant elliptical galaxy M86 has a total mass of about one tenth of the
whole cluster mass (Schindler et al. 1999).
The perturbation of the total gravitational field due to M86 is thus not
negligible. We have added the gravitational potential corresponding to M86
($\beta$=0.5, $r_{\rm C}$=74.2 kpc, and $\rho_{\rm C}=6.26\times 10^{-3}$ 
M$_{\odot}$pc$^{-3}$). First, we have made test
runs where we have put M86 on an elliptical orbit around M87.
The results for the orbiting perturber (M86) are that (i) M87 begins also to move 
and (ii) the additional mass distribution perturbs the trajectories of the 
test galaxy producing a wide variety of radial orbits with different impact
parameters.
Without the perturbing gravitational potential the orbits would lie
in one single plane and the impact parameter, i.e. the closest approach
to the cluster center, would stay constant.
Second, we have fixed M86 in space. In this case the galaxy orbits are perturbed
in the same way as in the previous case.
We can thus conclude that in both cases the presence of M86 near M87 
changes significantly the trajectories of galaxies on radial orbits. 
For simplicity we have chosen to fix M86 in space.

\begin{figure}
	\resizebox{\hsize}{!}{\includegraphics{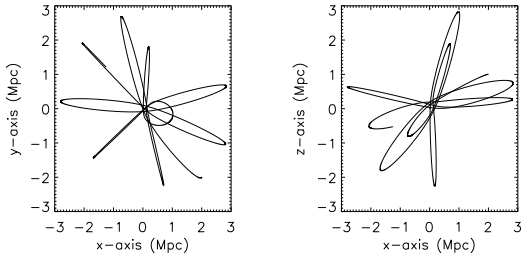}}
\figcaption[figure2.ps]{Test particle trajectory calculated with the fixed gravitational 
potential of M87 and M86. Left: $x$-$y$-plane. Right: $x$-$z$-plane.
This simulation does not trace the trajectory of one single galaxy but shows 
a variety of orbits corresponding to different initial conditions.
\label{fig:orbits}}
\end{figure}

Fig.~\ref{fig:orbits} shows the orbits
calculated with the fixed gravitational potential of M87 and M86. 
The asymmetric gravitational potential leads to a variation
of the impact parameter and the maximum velocity for each orbit.
The initial conditions were fixed such that the resulting
maximum ram pressure varies between 1000 and 
5000~cm$^{-3}$\,(km\,s$^{-1}$)$^{2}$. If a spiral galaxy is gravitationally
bound to the cluster, these high eccentricities are needed in order
to obtain the highest ram pressure maxima. However, 
for values of the maximum ram pressure smaller than
$\sim$3000~cm$^{-3}$\,(km\,s$^{-1}$)$^{2}$ slightly less eccentric orbits are
also possible.
The smallest impact parameter, i.e. the closest approach to the cluster center,
is $\sim$90~kpc with a maximum velocity of 2200~km\,s$^{-1}$.
A small change of the initial conditions, resulting in an impact parameter
of $\sim$70~kpc, leads to a maximum ram pressure of 10000~cm$^{-3}$\,(km\,s$^{-1}$)$^{2}$.
Given the uncertainties of the determination of the enclosed mass and the
ICM density profile, we think that this extreme orbit is still realistic
for the Virgo cluster.
The largest impact parameter is $\sim$200~kpc.
The timescale between two passages through the cluster core is about
10$^{10}$~yr. A galaxy will therefore have at most a few passages during
its lifetime within the cluster.
Thus the simulation does not trace the trajectory of one single galaxy but creates 
a variety of orbits corresponding to different initial conditions.
Nevertheless, it is not excluded that a galaxy can have two passages through
the cluster core. 

We define the normalized ram pressure as 
\begin{equation}
p_{\rm norm}=(\rho_{\rm ICM} v_{\rm galaxy}^{2})/(\rho_{0} v_{0}^{2})\ ,
\end{equation}
where $\rho_{0}$=10$^{-4}$ cm$^{-3}$ and $v_{0}$=1000 km\,s$^{-1}$
are representative averaged values. The resulting normalized ram pressure of the orbits 
shown in Fig.~\ref{fig:orbits} can be seen in Fig.~\ref{fig:normrampress}.
This graph illustrates that the perturbation of the galaxy orbits by
M86 can result in a change of the ram pressure strength up to 50\%
because of the small core radius of the ICM distribution.
A small change of the impact parameter leads to a considerable change
of the ICM density, whereas
for values of the impact parameters between 100 and 200~kpc the change of the
maximum galaxy velocities is only of the order of $\sim$10\%.     
For these radial orbits ram pressure dominates the
thermal pressure of the ICM ($T\sim 10^{7}$ K) for distances less
than one Abell radius ($\sim 2.5$~Mpc) as can be seen in Fig.~\ref{fig:Pram_PICM}.

In order to compare these orbits directly with observations,
we plot in Fig.~\ref{fig:orbits_vz} the line-of-sight (LOS) velocity
with respect to the cluster mean velocity as a function of the projected 
distance to the cluster center (M87).
The number of orbits is sufficient to be representative
for radial orbits over the whole volume of the Virgo cluster. The region
above the envelope of these curves corresponds to more circular orbits. 
This diagram will be used later on to compare the model orbits directly 
with observations.

We can now address the question if we can constrain the orbital shape of
H{\sc i} deficient galaxies. As stated above, a minimum ram pressure of
$p_{\rm ram}^{\rm min}=\rho_{\rm ICM}\,v_{\rm galaxy}^{2}
\sim 500$~cm$^{-3}$\,(km\,s$^{-1}$)$^{2}$ is necessary in order
to affect the atomic gas disk. In the case of circular orbits
the galaxy velocity has the Keplerian value 
$v_{\rm galaxy}=v_{\rm K}=\sqrt{R({\rm d}\Phi/{\rm d}R)}$, where
$\Phi$ is the gravitational potential of the cluster.
This means that in the Virgo cluster a galaxy with a rotation velocity of 
$v_{\rm rot}\sim 150$~km\,s$^{-1}$, which is on a 
circular orbit, must be closer than 250~kpc (50$'$) to the cluster center
in order to be affected significantly by ram pressure. 
If spiral galaxies have been accreted
only recently, it is not probable that they are on circular orbits with a such
small distance to the cluster center.
On the observational side, Hoffman, Helou, \& Salpeter (1988) found a few H{\sc i}
rich dwarf galaxies within a radius of 5$^{\rm o}$ around M87. These galaxies are much
more vulnerable to ram pressure stripping, because of their lower gravitational
potential. At least some of these dwarfs must evolve on circular orbits with 
radii smaller than 1~Mpc, where they have never been stripped by ram pressure.
 
In the case of highly eccentric orbits, the maximum orbital velocity is given by the
escape velocity $v_{\rm galaxy}\sim \sqrt{2}\,v_{\rm K}=v_{\rm escape}$.
The impact parameter must be smaller than
400~kpc (1.3$^{\rm o}$) in order to obtain $p_{\rm ram} >  500$~cm$^{-3}$\,(km\,s$^{-1}$)$^{2}$. 
For more massive galaxies ($v_{\rm rot}>150$~km\,s$^{-1}$) the 
maximum impact parameter is still smaller. The minimum impact parameter of 90~kpc excludes 
a gravitational interaction of the galaxy with M87.

The impact parameters of the orbits shown in Fig.~\ref{fig:orbits} are in the range 
between 90 and 200~kpc giving rise to maximum ram pressure 
$p_{\rm ram}=$1000--5000~cm$^{-3}$\,(km\,s$^{-1}$)$^{2}$. 
We will show that galaxies on these orbits will have an H{\sc i} deficiency 
$DEF > 0.3$ after a passage through the cluster center.

\section{The model}

There are different ways to model the interstellar gas numerically:
(i) hydrodynamical, (ii) SPH, and (iii) sticky particle models.
Each class has its advantages and limits. Since the ISM is neither a
continuous medium (hydrodynamics, SPH) nor exclusively made of clouds
(sticky particles), one has to chose the class which is well adapted for
the investigated astrophysical problem. In the following we will motivate the
choice of our model.

The interstellar H{\sc i} gas is a clumpy medium
(see e.g. Kim et al. 1998 for the LMC). In the Galaxy roughly half of the
interstellar H{\sc i} is in the warm phase ($T\sim 6000$~K) with a
volume filling factor of $\sim$0.5. The other half is in the form
of cold ($T\sim$80~K) and dense clouds with a filling factor of $\sim$0.01
(Kulkarni \& Heiles 1988).
If they are in pressure equilibrium with the rest of the ISM at $nT=3000$~K\,cm$^{-3}$,
they have a density of $n\sim$40~cm$^{-3}$.
The largest clouds have column densities of $\sim 2 \times 10^{20}$~cm$^{-2}$.
On the largest scale, three dimensional structures
of the atomic gas phase are major H{\sc i} concentrations around dark clouds and newly 
formed stars. They have column densities of $N_{\rm HI} \sim 10^{21}$~cm$^{-3}$ and masses
of $\sim 10^{5}$~M$_{\odot}$ (Kulkarni \& Heiles 1988). 
Thus, the cold atomic phase of the ISM is well represented by discrete particles.
Concerning the warm neutral phase, Murakami \& Babul (1999) have shown in their
hydrodynamical simulations of the effects of ram pressure on an
elliptical galaxy that the continuous ISM tends to be stripped in the form of distinct clouds if they included the external ICM pressure.
These clouds have densities $n \sim 1$~cm$^{-3}$ and radii $R_{\rm cl} \sim 300$~pc, 
resulting in a column density of $N_{\rm HI} \sim 10^{21}$~cm$^{-3}$. 
We therefore think that the warm neutral phase, which is subjected to 
an external pressure due to the surrounding ICM and to ram pressure forces,
can also be modeled by distinct particles.

These considerations motivate our approach to model the entire atomic phase of the
ISM as sticky particles where each particle represents a gas cloud complex.
We assume that the clouds have a constant column density.
The relevant physical property of the ISM with respect to ram pressure is the 
column density of the gas/clouds. It is adapted to observed values.

The effect of ram pressure is simulated as an additional acceleration applied 
on the clouds which are exposed to the ICM in the direction of the
galaxy's motion within the cluster.

\subsection{\label{sec:galmodel} The model galaxy} 

Since we do not want to include gravitational interactions, we
let the gas clouds evolve in a given analytical gravitational
potential. It consists of two spherical components and a disk
(Allen \& Santill\'an 1991):
\begin{itemize}
\item
the dark matter halo potential which is given by 
\begin{equation}
\Phi_{\rm halo}(R)=-(\frac{M(R)}{R})-(\frac{M_{\rm halo}}{1.02a_{\rm halo}})\times
\end{equation}
\begin{equation}
\big[ -\frac{1.02}{1+(R/a_{\rm halo})^{1.02}}+{\rm ln}(1+(R/a_{\rm halo})^{1.02})
\big]^{100}_{R}\ ,
\end{equation} \label{eq:halopotential}
where 
\begin{equation}
M(R)=\frac{M_{\rm halo}(R/a_{\rm halo})^{2.02}}{1+(R/a_{\rm halo})^{1.02}}.
\end{equation} \label{eq:halopotentialmass}
R is the distance to the galaxy center, $M_{\rm halo}=8.6\times 10^{10}$ 
M$_{\odot}$, and $a_{\rm halo}$=12 kpc.
\item
The bulge potential which is given by
\begin{equation}
\Phi_{\rm bulge}(R)=-\frac{M_{\rm bulge}}{\sqrt{R^{2}+b_{\rm bulge}^{2}}}\ ,
\end{equation} \label{eq:bulgepotential}
where $M_{\rm bulge}=5.6\times 10^{9}$ M$_{\odot}$ and $b_{\rm bulge}$=387 pc.
\item
The disk potential which is given by
\begin{equation}
\Phi_{\rm disk}(r,z)=-\frac{M_{\rm disk}}{\sqrt{r^{2}+(a_{\rm disk}+\sqrt{z^{2}+b_{\rm disk}^{2}})^{2}}}\ ,
\end{equation} \label{eq:diskpotential}
where $R^{2}=r^{2}+z^{2}$, $M_{\rm disk}$=2.6$\times 10^{10}$~M$_{\odot}$, 
$a_{\rm disk}$=2.7~kpc, and $b_{\rm disk}$=250~pc.
\end{itemize}
We fix the optical diameter of the disk at $D_{\rm opt}=$20~kpc
($\sim$7.5 disk scale lengths).
We have included selfgravity between the cloud complexes.
Fig.~\ref{fig:initrotationvel} shows the resulting rotation curves 
after a disk evolution of 10$^{9}$ yr for the different components as well 
as the decomposition and the total model rotation curve. The latter is used as 
initial conditions for our simulations. We thus obtained a constant rotation curve 
of $\sim$140~km\,s$^{-1}$. This rotation velocity corresponds to a medium
mass spiral galaxy.
Fig.~\ref{fig:azimuthal} shows the H{\sc i} cloud distribution in
$z$ direction after 10$^{9}$ yr. We end up with a realistic height
of the H{\sc i} layer of $\sim$1000 pc at a radius of 10 kpc. 
The initial H{\sc i} diameter is $D_{\rm HI}^{\rm init}$=30~kpc.

The cloud mass distribution is $n(m)\propto m^{-1.5}$ (Knude 1981).
The initial total gas mass of the galaxy is $6\times 10^{9}$ M$_{\odot}$,
the total number of clouds is 10\,000. 
We adopt a constant H{\sc i} column density for all clouds
(Sanders, Scoville \& Solomon 1985) which is fixed at $\Sigma_{\rm cl}=
7.5 \times 10^{20}$ cm$^{-2}$. This value is consistent with
the measurements made by Rots et al. (1990) on nearby galaxies.
Assuming the surface density to be constant has the advantage that
the acceleration due to ram pressure is the same for all clouds
disregarding their masses. This leads to the following  mass-radius relation
which is relevant for the collision rate:
\begin{equation}
r_{\rm cloud}^{\rm coll}=\sqrt{M_{\rm cloud}/(\pi\,\lambda\,\Sigma)}\ ,
\end{equation} \label{eq:mass_radius}
where $\Sigma$ is the gas surface density of one cloud. In reality the
number of H{\sc i} clouds is much larger than the number of particles
in our model. To compensate this effect we have chosen the factor $\lambda$
in the way that the mean free path of one cloud is the same
for 10\,000 clouds as for a realistic cloud number.

In addition, in the interior of the galaxy gas clouds become denser building 
molecular cores. This results in smaller cloud sizes and higher surface 
densities. As the atomic gas represents the outer layer of these clouds,
they can be hardly pushed by the ram pressure. We take this effect 
into account by adding a factor $\gamma$ in the equation for the acceleration
due to ram pressure:
\begin{equation}
{\bf a}_{\rm ram}=p_{\rm ram}/(\gamma \,\Sigma_{\rm cl})\ ,\ \ 
\gamma=15\,\exp\big(-(r/r_{0})\big)+1
\end{equation}
where $r_{0}$=2~kpc $(\simeq 0.4''$ for the Virgo cluster). 
This profile is consistent with the CO(1--0) observations of Virgo cluster 
spirals (Kenney \& Young 1988).

When orbiting around the galaxy center, the clouds can have
inelastic collisions. During these collisions they can exchange mass
and larger clouds can grow through coalescence. The outcome of
a collision can be one (coalescence), two (mass exchange), or three 
fragments (fragmentation).
Let the radius of the first cloud be $r_{1}$, that of the second
cloud $r_{2}$. Let the impact parameter be $b$, the velocity of
the fragment $v_{\rm f}$, and the escape velocity $v_{\rm esc}$. 
We follow the prescriptions of Wiegel (1994):
\begin{itemize}
\item
for $r_{1}-r_{2} < b < r_{1}+r_{2}$:\\ fragmentation
\item
for $b \le r_{1}-r_{2}$ and $v_{\rm esc} > v_{\rm f}$:\\ 
mass exchange
\item
for $b \le r_{1}-r_{2}$ and $v_{\rm esc} \le v_{\rm f}$:
\\ coalescence
\end{itemize}
This description is similar to that of Combes \& G\'erin (1985) 
or Klari\'c (1995). The search for the next neighbors is done with 
the help of a treecode (Barnes \& Hut 1986).
The integration of the ordinary differential equation is done with the
Burlisch-Stoer method (Stoer \& Burlisch 1980) using a Richardson 
extrapolation and Stoermer's rule.
This method advances a vector of dependent variables $y(x)$ from a point
$x$ to a point $x+H$ by a sequence of $n$ substeps.
Thus, the initial timestep $H$ is divided subsequently into 
$n$=2, 3, 4, etc. substeps. At the end the solution
of $y(x+H)$ is extrapolated and an error can be estimated. 
The size of the timestep is adaptive and linked to the estimated
error of the extrapolation. This error is normalized by the 
values of the distance covered during the last timestep and the velocity
of each particle. The error level for acceptance of the extrapolated
solution is a free parameter and has to be adapted to the
physical problem treated. For a relative error level $\epsilon < 0.1$
the Courant criterion is fulfilled for each timestep.\\
The collisions are evaluated at each timestep $h=H/n$ and only those
which appear for all sequences $n$ are taken into account.
We adopt the strategy to chose the error level in a way to have timesteps 
of the order of several 10$^{4}$ yr. 

Ram pressure is included as an additional acceleration of those
clouds which are exposed to the ICM in the direction of the galaxy motion.
Fig.~\ref{fig:windy} illustrates this effect:
clouds located inside the disk are protected by clouds on the surface due
to their finite size (this depends
of course strongly on the inclination angle of the galaxy
with respect to the orbital plane). 
As it can be seen in Fig.~\ref{fig:orbits} the galaxy trajectories
near the cluster center are quasi linear. This means that we
do not have to change the wind direction during the simulation. 

\section{The simulations}

\begin{figure}
	\resizebox{\hsize}{!}{\includegraphics{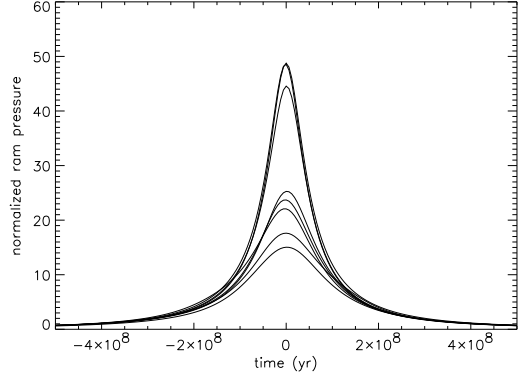}}
\figcaption[figure3.ps]{Normalized ram pressure (in units of $\rho_{0} v_{0}^{2}$) 
as a function of time for the galaxy orbits shown in Fig.~\ref{fig:orbits}. 
\label{fig:normrampress}}
\end{figure}

\begin{figure}
	\resizebox{\hsize}{!}{\includegraphics{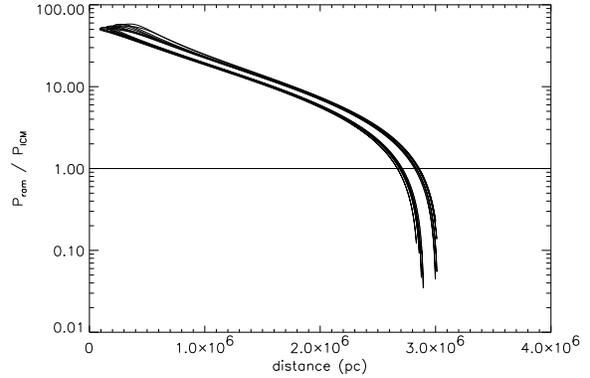}}
\figcaption[figure4.ps]{Ratio of the ram pressure and thermal pressure for 
the galaxy orbits shown in Fig.~\ref{fig:orbits} as a function of the
projected distance to the cluster center. \label{fig:Pram_PICM}}
\end{figure}

\begin{figure}
	\resizebox{\hsize}{!}{\includegraphics{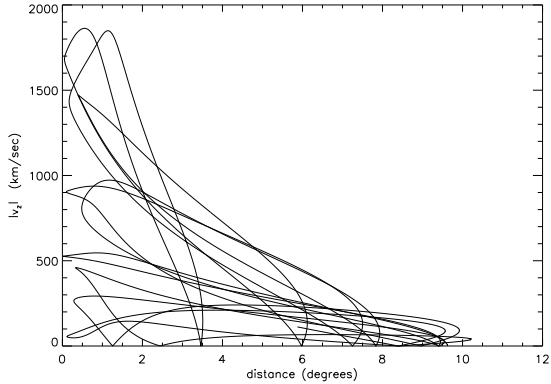}}
\figcaption[figure5.ps]{Line-of-sight (LOS) velocity with respect to the 
cluster mean velocity as a function of the projected distance to the cluster 
center for the galaxy orbits shown in Fig.~\ref{fig:orbits}. 
\label{fig:orbits_vz}}
\end{figure}

\begin{figure}
	\resizebox{\hsize}{!}{\includegraphics{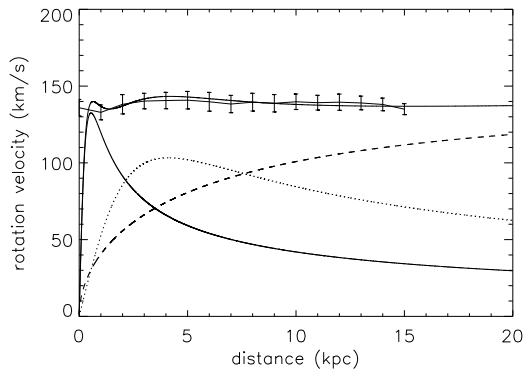}}
\figcaption[figure6.ps]{Decomposition of the rotation curve after a disk
evolution of 10$^{9}$~yr. 
Dashed: halo, solid: bulge, dotted: disk; thick: total, solid with error bars: 
initial model rotation curve. The error bars represent the velocity 
dispersion of the N--body model. \label{fig:initrotationvel}}
\end{figure}

\begin{figure}
	\resizebox{\hsize}{!}{\includegraphics{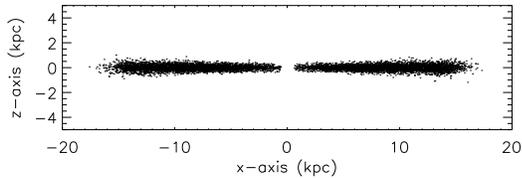}}
\figcaption[figure7.ps]{H{\sc i} cloud distribution in $z$ direction 
after a disk evolution 10$^{9}$~yr. \label{fig:azimuthal}}
\end{figure}

\begin{figure}
	\resizebox{\hsize}{!}{\includegraphics{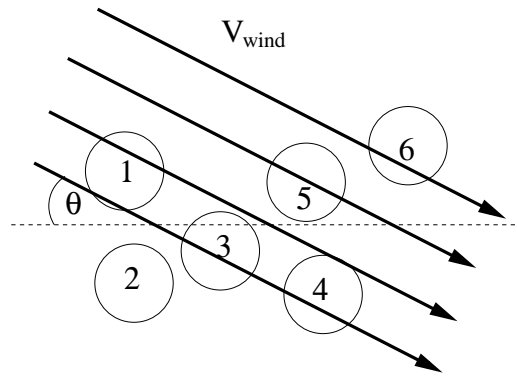}}
\figcaption[figure8.ps]{Illustration of the effect of ram pressure. The 
direction of the wind created by the galaxy's motion in the ICM is indicated 
by the arrows. Only clouds which are directly exposed to the wind are pushed 
by ram pressure (clouds no.1, 2, 5, and 6). The dashed line represents the
galactic plane, $\Theta$ is the inclination between the galactic plane and the 
wind direction. \label{fig:windy}}
\end{figure}

\begin{figure}
	\resizebox{\hsize}{!}{\includegraphics{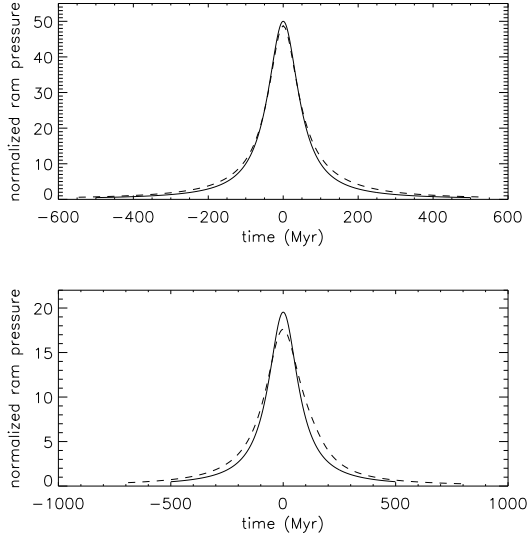}}
\figcaption[figure9.ps]{Dashed: ram pressure profiles of Fig.~\ref{fig:normrampress}. 
Solid: analytical ram pressure profile (Lorentzian, see Eq.~\ref{eq:lorentzian}).
\label{fig:rampressureprofile}}
\end{figure}

Fig.~\ref{fig:rampressureprofile} shows the profile of two ram pressure 
stripping events of Fig.~\ref{fig:normrampress} as a dashed line. In our simulations
we approximated this ram pressure profile with a Lorentzian profile
(Fig.~\ref{fig:rampressureprofile} solid line): 
\begin{equation}
p_{\rm ram}= p_{\rm max} \times (t^{2}+t_{\rm HW}^{2})^{-1}\ ,
\label{eq:lorentzian}
\end{equation}
where $p_{\rm max}$ is the maximum ram pressure at the smallest distance to
the cluster center and $t_{\rm HW}$ is the duration of the event.
The simulations were started at $t$=-500~Myr. 
For a fixed $p_{\rm max}$, $t_{\rm HW}$ is chosen in order to obtain a constant
ram pressure at $t$=-500~Myr: $p_{\rm ram}(-500\,{\rm Myr})=0.5\rho_{0}v_{0}^{2}$. 
This procedure reproduces best the ram pressure profiles of the model
orbits (Fig.~\ref{fig:normrampress}).

These values together with the wind direction
and the resulting final parameters for each run can be seen in 
Table~\ref{tab:runs}. The columns are the following: Col. (1): RUN. 
Col. (2): Maximum ram pressure in units of $\rho_{0}v_{0}^{2}$. 
Col. (3): Inclination angle $\Theta$ with respect to the orbital plane in degrees.
We want to recall here that $\Theta$=0$^{\rm o}$ means edge-on stripping
and $\Theta$=90$^{\rm o}$ face-on stripping.
Col. (4): Time between the beginning of the stripping (defined by the moment when 
the total gas mass within 25 kpc $M_{\rm tot} < 5\times10^{9}$ M$_{\odot}$) and the
galaxy's closest passage to the cluster center.
Col. (5): Radial distance corresponding to Col. (4) assuming a constant
velocity of the galaxy with respect to the cluster center of 
$v_{\rm gal}$=1700 km\,s$^{-1}$. 
Col. (6): Final H{\sc i} diameter in kpc. Col. (7): Final central surface 
density in M$_{\odot}$\,pc$^{-2}$. Col(8): Final H{\sc i} deficiency 
(see Sect.~\ref{sec:totmass}).
Col.(9): Re-accreted gas mass in $10^{8}$~M$_{\odot}$.
Col.(10): Re-accreted gas mass divided by the stripped gas mass.
The final H{\sc i} diameter is defined by a limiting column density
of $10^{20}$~cm$^{-2}$. Following Cayatte et al. (1994), the radial profile
was determined by averaging the model distribution in rings of integrated column density.
The central surface density is defined as the mean density over the inner
part of the disk out to a quarter of the optical radius. For the calculation of the
stripped and the re-accreted gas mass we used the enclosed mass within a radius
of 20~kpc and a height of 1~kpc. For RUN A(10,0) $2.4 \times 10^{8}$~M$_{\odot}$
of atomic gas are pushed to radii greater than 20~kpc and then fall
back within 400~Myr. Thus, the ratio between stripped and re-accreted gas is
greater than one.

\begin{figure}
	\resizebox{\hsize}{!}{\includegraphics{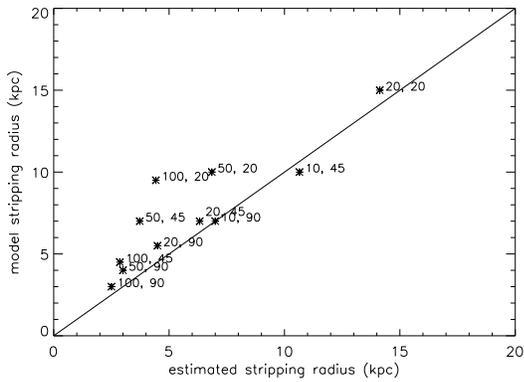}}
\figcaption[figure10.ps]{Final H{\sc i} radius $R_{\rm HI}=D_{\rm HI}/2$ 
of our simulations as a function of the estimated radius $R_{\rm GG}$ using
the formula of Gunn \& Gott (1972). The solid line corresponds to 
$R_{\rm GG}=R_{\rm HI}$. \label{fig:gunn_gott}}
\end{figure}

We have made 16 runs using different disk inclinations with respect
to the orbital plane and different values for the maximum ram pressure. 
We only use values for the maximum ram pressure which are high enough to 
show a visible deformation of the H{\sc i} distribution.
In Fig.~\ref{fig:gunn_gott} we show the final H{\sc i} radius $R_{\rm HI}=D_{\rm HI}/2$ 
of our simulations as a function of the linearized estimated radius 
using the formula of Gunn \& Gott (1972):
$R_{\rm GG}=\Sigma_{\rm gas}v_{\rm rot}^{2}/(\rho_{\rm ICM}v_{\rm galaxy}^{2})+2$~kpc,
where $\Sigma_{\rm gas}=7.5 \times 10^{20}$~cm$^{-2}$ and $v_{\rm rot}=140$~km\,s$^{-1}$.
The additional length is due to the length scale of the molecular gas 
distribution ($\gamma$). 
The solid line corresponds to $R_{\rm GG}=R_{\rm HI}$.
As expected, the estimated
stripping radius using Gunn \& Gott's formula is well reproduced
for an inclination angle $\Theta =90^{\rm o}$ (face-on stripping) . 

In order to generalize Gunn \& Gott's formula for inclination angles $\Theta \neq 90^{\rm o}$, 
we searched for a simple correlation between 
the stripping radius and ($\sin \Theta$, $\rho v^{2}$). Since we did not succeed,
we conclude that it can be dangerous to
derive the stripping parameters only on the basis of the final H{\sc i} diameter.
In particular, taking the velocity component perpendicular to the disk plane
$p_{\rm ram}=\rho v_{\perp}^{2}$ does not reproduce the final H{\sc i} radii
of our simulations. We will show in Section~\ref{sec:amount} that the H{\sc i} deficiency 
is much better suited.

\begin{figure}
	\resizebox{\hsize}{!}{\includegraphics{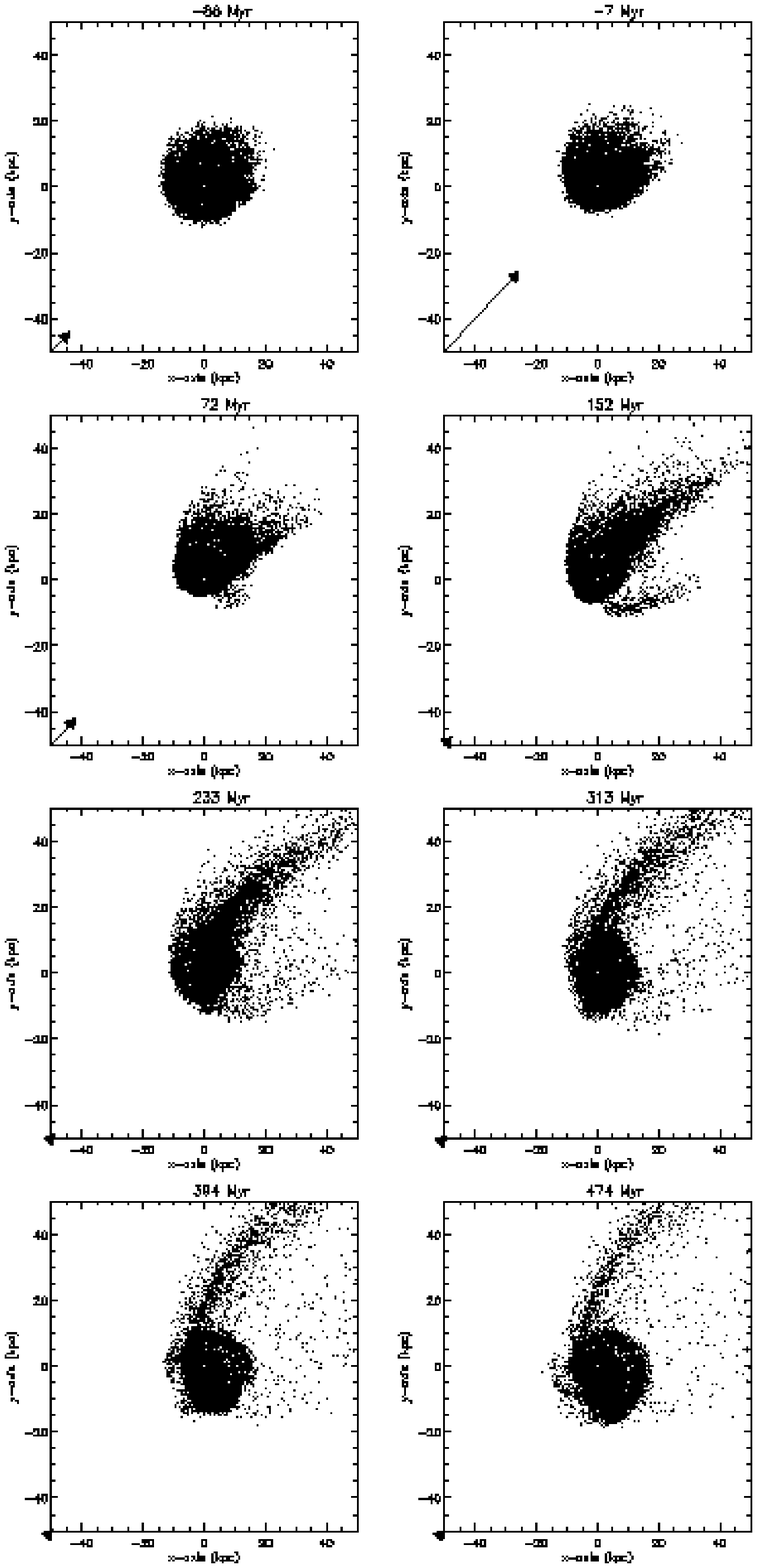}}
\figcaption[figure11.ps]{Snapshots of RUN C(50, 0). The elapsed time is 
indicated at the top of each panel. The galaxy is seen face-on and is moving 
in the south-east direction, i.e. the wind is coming from the south-east. 
This wind direction is indicated by the arrows whose length is proportional
to $\rho v^{2}$. \label{fig:simulation1}}
\end{figure}

In order to give an idea of how ram pressure acts in detail, 
Fig.~\ref{fig:simulation1} shows an example for an edge-on stripping process
(RUN C(50,0)). The ram pressure maximum occurs at t=0~yr.
The wind direction is indicated by the arrows whose length is proportional
to ram pressure.
At $t\sim -100$~Myr, a density enhancement begins to grow in the 
directions of the galaxy's motion. At $t \sim$70~Myr, the gas whose 
rotation velocity is parallel to the wind direction is accelerated and
driven out of the
galaxy leading to an arm-like structure parallel to the wind direction. 
At the opposite side where the gas clouds are decelerated by the wind,
a second much less pronounced arm forms. At $t$=150~Myr ram
pressure has completely ceased and the dynamics are determined by rotation
and re-accretion. The main arm moves to the north-west
of the galaxy due to its initial angular momentum. At the same time,
the gas begins to fall back to the galaxy in the north and south-west.
This infalling gas is forming a second arm in the south-west. 
At $t \sim$300~Myr a density enhancement in the east is observed which is due
to the infalling material from the north.
It is interesting to notice that the gas within the disk of the galaxy 
forms a leading m=1 spiral structure. 

\begin{figure}
	\resizebox{\hsize}{!}{\includegraphics{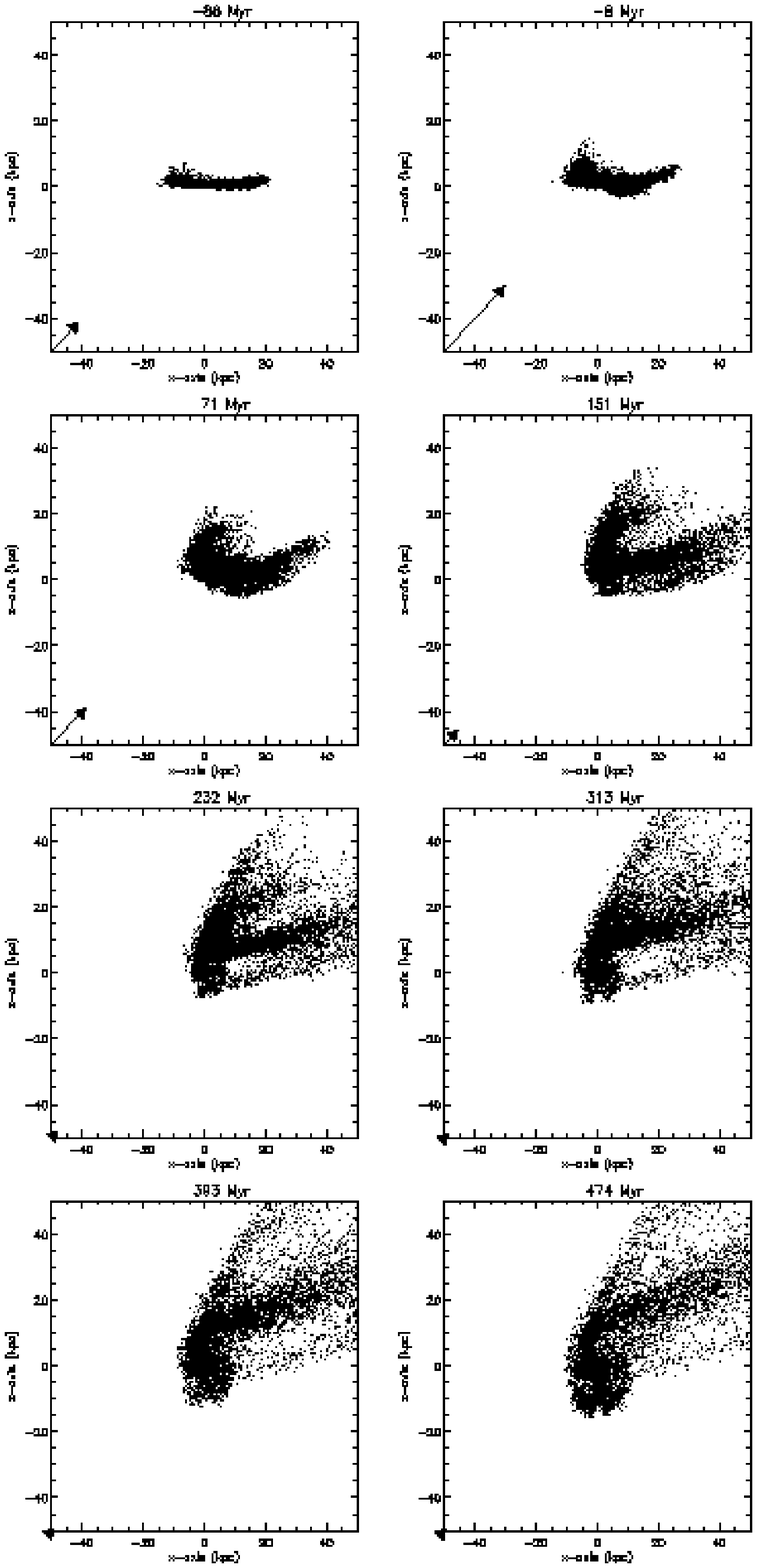}}
\figcaption[figure12.ps]{Snapshots of RUN J(20, 45). The elapsed time is 
indicated at the top of each panel. The galaxy is seen edge-on and is moving 
in the south-east direction, i.e. the wind is coming from the south-east. 
This wind direction is indicated by the arrows whose length is proportional
to $\rho v^{2}$. \label{fig:simulation}}
\end{figure}

We show snapshots of RUN J(20,45) in Fig.~\ref{fig:simulation} as an
example of an out of plane stripping beginning at with time differences
of $\Delta t\sim 80$~Myr. The galaxy is seen edge-on and 
is moving in the south-east
direction, i.e. the wind is coming from the south-east. This wind
direction is indicated by the arrows whose length is proportional
to ram pressure. The ram pressure maximum occurs at $t$=0~yr.
At $t\sim$-90~Myr the disk is already slightly deformed in $z$-direction.
When ram pressure is at maximum the outer parts of the gas disk are pushed
in the wind direction. Ram pressure exceeds the threshold of
$p_{\rm ram}= 5 \rho_{0} v_{0}^{2}$ early ($t \sim$-300~Myr) for the given
Lorentzian profile. An important distortion of the gas distribution 
can be observed at $t \sim$0~yr. The gas is driven out of the galaxy
in the direction of the wind and rotates at the same time.
This creates the western arm which develops between $t$=-10~Myr and 
$t$=100~Myr.  
Since the duration of the ram pressure event ($t_{\rm HW}$) is a
considerable fraction of the rotation period, the out of plane gas
in the region where its motion due to rotation and the motion of the ICM
are opposite is compressed. This compressed region corresponds to
the arm which develops in the north of the galaxy. For $t >$200~Myr
the wind has completely ceased and the evolution of the gas distribution
is due to rotation and re-accretion of the stripped gas.
At the end of the simulation ($t$=500~Myr) a considerable part
of the stripped ISM is located within a radius of 15~kpc outside the disk.
Even at $t \sim$500~Myr after the galaxy's closest passage to the cluster center 
the column density of the gas located outside
the galactic disk is of the order of 10$^{20}$~cm$^{-2}$.

\section{The fate of the stripped gas \label{sec:fate}}

We now address the question if such a large structure can be observed and
in which phase the stripped gas is re-accreted.
Murakami \& Babul (1999) found that, if the thermal gas pressure of the ICM 
is taken into account, massive (10$^{5}$-10$^{6}$ M$_{\odot}$)
neutral gas clouds are stripped. They speculate that they might be able to 
survive in the hot ICM. Can an H{\sc i} blob survive long enough 
(i.e. during several 10$^{8}$~yr) in a hostile environment like the ICM? 
In order to estimate the evaporation time of a stripped gas cloud, we follow
Cowie \& McKee (1977). The classical approximation for the evaporation
rate breaks down when the electron mean free path becomes comparable or 
larger than the temperature scale height. In this case the classical heat flux
exceeds the saturated heat flux and
\begin{equation} \label{eq:sigma}
\sigma_{0} \simeq \big( \frac{T_{\rm ICM}}{1.54\times 10^{7} {\rm K}}\big)^{2} 
\frac{1}{n_{\rm ICM}R_{\rm pc}}>1
\end{equation}
where $T_{\rm ICM}$ is the temperature of the ICM, $n_{\rm ICM}$ is the
ICM density, and $R_{\rm pc}$ is the cloud radius.
Assuming  $R_{\rm pc}$=10 pc, $T_{\rm ICM}$=10$^{7}$ K, and 
$n_{\rm ICM}$=10$^{-4}$ cm$^{-3}$ leads to $\sigma_{0}$=422.
The saturated evaporation time is given by
\begin{equation} \label{eq:tevap}
t_{\rm evap}^{\rm sat}=2.8\times 10^{6}(n_{\rm c}/n_{\rm ICM})R_{\rm pc}
/(T_{\rm ICM}^{1/2}\,2.73\sigma_{0}^{3/8})\ {\rm yr}
\end{equation}
For a given cloud column density $N_{\rm c}$ this gives 
$t_{\rm evap}^{\rm sat}\simeq 10^{7} (N_{\rm c}/10^{20}\ {\rm cm}^{-2})$ yr.
With an initial column density of $N_{\rm c}=10^{21}$~cm$^{-2}$, a stripped cloud evaporates
within 10$^{8}$~yr. In the case of a magnetic field configuration which
inhibits heat flux (e.g. a tangled magnetic field), this evaporation time can increase
significantly (Cowie, McKee, \& Ostriker 1981) and might attain several 10$^{8}$~yr.

On the other hand, if a stripped cloud leaves the galaxy, its main heating source,
the stellar FUV radiation field, disappears. The thermal equilibrium gas temperature is then
determined by heating due to the soft X-ray emission of the ICM and cooling due to the 
infrared line emission of C{\sc ii} and O{\sc i}. 
Thus, when the cloud is stripped, cooling dominates heating, the gas
cools down, becomes denser and begins to form H$_{2}$ within a timescale of
$\tau_{{\rm H}_{2}}\sim 10^{9}/n_{\rm c}$~yr (Hollenbach \& Tielens 1997).
The only atomic hydrogen resides in the outer part of the clouds where
the X-ray flux dissociates the molecular hydrogen. 
We will now estimate the column density of this H{\sc i} layer.

Maloney, Hollenbach, \& Tielens (1996) have modeled the physical and chemical state
of dense neutral gas exposed to an external X-ray source.
We will extrapolate their results, which they have obtained for higher gas densities
and X-ray fluxes. These authors assumed a power low X-ray spectrum,
whereas in our calculations the radiation field is made of thermal bremsstrahlung. 
Despite this difference we can use their results, because for hot gas temperatures of 
order 10$^{7}$~K, the normalization constant for a given X-ray 
flux is not very different from that for a spectral index $\alpha=0.7$
adopted by Maloney et al. (1996). In order to apply their calculations
we have to make sure that photons with energies less than 100 eV do not 
contribute to the ionization of the molecular cloud. This translates to the
condition that the ionized column density due to photons of $E<100$~eV
must be much smaller than the total column density.

The ionizing photon flux is given by
the integral of d$E/E$ over the range from 13.6~eV to 100~eV:
\begin{equation}
\phi_{\rm i}=\frac{\ln (0.1/0.0136)}{1.6 \times 10^{-9}} \frac{F_{\rm X}}{1.5}=
8.3 \times 10^{8} F_{\rm X}\ {\rm photons\,cm}^{-2}{\rm s}^{-1}\ ,
\end{equation}
where $F_{\rm X}=(2-0.5) F_{0}$ is the X-ray flux in the ROSAT band.
In equilibrium this gives rise to an ionized column density
\begin{equation}
N_{\rm e}=\frac{\phi_{\rm i}}{\alpha n_{\rm e}}\ ,
\end{equation}
where $\alpha=2.6 \times 10^{-13} (T/10^{4}\ {\rm K})^{-0.8}$~cm$^{3}$s$^{-1}$ is
the recombination coefficient and $n_{\rm e}$ is the electron density.
Using $F_{\rm X}=1.5 \times 10^{-5}$~erg\,cm$^{-2}$s$^{-1}$ (see below), $T=2 \times 10^{7}$~K,
and $n_{\rm e}=100$~cm$^{-3}$ gives $N_{\rm e} \sim 10^{17}$~cm$^{-2}$.
This is much smaller than the total cloud density of $\sim 10^{20}$~cm$^{-2}$.

Maloney et al. (1996) showed that the transition between atomic and molecular hydrogen 
occurs if $H_{\rm X}/n_{\rm c} \geq 10^{-28}$~ergs\,cm$^{3}$\,s$^{-1}$,
where $H_{\rm X}$ is the energy deposition rate per particle and $n_{\rm c}$ is the
local cloud density. This deposition rate is approximately 
\begin{equation}
H_{\rm X} \sim 5.8\times 10^{-23}\,F_{\rm X} \big(\frac{N_{\rm H}}{10^{22}\,{\rm cm}^{-2}}\big)\ {\rm ergs}\,{\rm s}^{-1}\ ,
\end{equation}
where $N_{\rm H}$ is the hydrogen column density.
In order to estimate the X-ray flux, we take the X-ray surface density profile given 
by Schindler et al. (1999)
\begin{equation}
F=F_{0}\big(1+(r/r_{\rm core})^{2}\big)^{-1}\ ,
\end{equation}
with a core radius $r_{\rm core}$=13.35~kpc. $F_{0}$ is determined by 
$2\pi \int_{0}^{r_{\rm out}} rS\,dr=L_{\rm X}$, where $r_{\rm out}$=1.5~Mpc is the outer  
radius and $L_{\rm X}=8.3\times 10^{43}$~ergs\,s$^{-1}$ is the total X-ray
luminosity (B\"{o}hringer et al. 1994). This gives
$F_{0}=1.15\times 10^{-3}$~ergs\,cm$^{-2}$\,s$^{-1}$. At a distance of $r$=0.15~Mpc 
from the cluster center the X-ray flux is 
$F\simeq 1.5\times 10^{-5}$~ergs\,cm$^{-2}$\,s$^{-1}$. 
With a hydrogen density of $n=100$~cm$^{-3}$ the H/H$_{2}$ transition thus occurs 
at a hydrogen column density of $N_{\rm H}\sim 2\times 10^{20}$~cm$^{-2}$. 
If the neutral gas is in pressure equilibrium with the ICM ($T_{\rm ICM}=10^{7}$~K, 
$n_{\rm ICM}\sim 10^{-3}$~cm$^{-3}$ at $r$=0.15~Mpc), it has a temperature of 
$T\sim 100$~K. 

The critical mass of an isothermal sphere (see e.g. Spitzer 1978)
with an external pressure at the boundary
of $p_{\rm ext}/k_{\rm B}=n_{\rm ICM}T_{\rm ICM}=10^{4}$~cm$^{-3}$K is 
$M_{\rm crit}=1.5\times 10^{3}$~M$_{\odot}$. If the magnetic field is of the order of a 
$B\sim 5\mu$G, the critical mass
might be as large as  $M_{\rm crit}\sim 5\times 10^{3}$~M$_{\odot}$.
If the temperature of the neutral gas is $T$=200~K this critical mass becomes
$M_{\rm crit}\sim 10^{4}$~M$_{\odot}$.

Such a cloud has a size of $\sim 10$~pc and a total column density of 
$N_{\rm tot}\sim 3 \times 10^{21}$~cm$^{-2}$.
Thus, only 10\% of the total column density of these clouds is in atomic form visible
in the H{\sc i} 21\,cm line. The total column density of the stripped gas in 
Fig.~\ref{fig:simulation} observed with a 20$''$ beam (VLA C+D array configuration 
at 21~cm) is $N_{\rm tot}^{\rm stripped}\sim {\rm several}\ 10^{20}$~cm$^{-2}$.
Without evaporation the observed H{\sc i} column density would thus be 
$N_{\rm HI}\sim {\rm several}\ 10^{19}$~cm$^{-1}$. 
High sensitivity H{\sc i} 21cm observations of 
Virgo spiral galaxies with a distorted atomic gas distribution
(see e.g. Phookun \& Mundy 1995) have shown that low surface brightness structures
appear when the sensitivity reaches several 10$^{19}$~cm$^{-2}$. 
This could be also the case for mildly H{\sc i} deficient galaxies $DEF < 0.4$,
where re-accretion is important (see Table~\ref{tab:runs}).

We suggest that most of the stripped gas, which can be found beyond the galactic disk, 
is thus very likely either hot ($T\geq 10^{6}$~K), ionized, and has a low density 
($n \geq n_{\rm ICM}$),
or molecular, cold ($T\sim 10-100$~K) and has a high density ($n \geq 100$~cm$^{-3}$).
In the second case, only $\sim$10\% of the total gas column density is in atomic
form due to X-ray dissociation.  

Combined H{\sc i}, CO, H$\alpha$, and X-ray observations will give informations
about the different phases of the ISM.
High resolution interferometric observations of single cluster galaxies are necessary because
they can be directly compared with our model snapshots. 
The detailed comparison between simulations and observations will be a crucial test for our model.
It will allow to draw conclusions
about the possible phase transitions (see above) of the ISM during the ICM--ISM interaction. 
We want to stress here that it is very important to compare both,
the gas distribution {\it and} the velocity field in order to have as much constraints as 
possible. First results were already obtained with this technique:
For the Virgo cluster spiral galaxy NGC~4522 Vollmer et al. (2000) succeeded in explaining 
extra planar H$\alpha$ emission
with a ram pressure scenario using one of the simulation snapshots. 
Vollmer et al. (2001) compared multi wavelength observations of the Coma cluster
galaxy NGC~4848 with a simulation snapshot.
They showed that all observations are consistent with the ram pressure scenario.
Studies of other cluster galaxies are in progress. The observations
of single cluster galaxies will show the way how to model the properties of 
the ISM (e.g. phase transitions, ionization, temperature) and their evolution
during an ISM--ICM interaction.

\section{Results}

\begin{figure}
	\resizebox{\hsize}{!}{\includegraphics{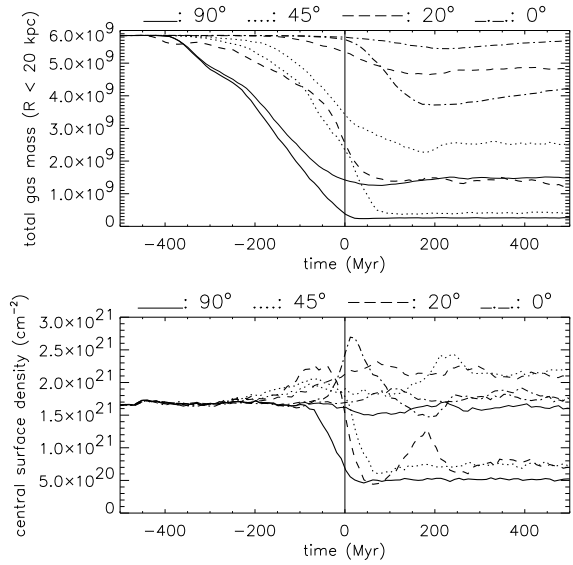}}
\figcaption[figure13.ps]{Variation of different galaxy parameters versus time. 
Upper panel: Total gas mass within 20~kpc $M_{\rm tot}$ in M$_{\odot}$.
Lower panel: Central surface density $\Sigma_{\rm central}$ in cm$^{-2}$.
In each panel two graphs are shown for the same inclination angles 
between the disk and the orbital plane corresponding to
$p_{\rm ram}=100 \rho_{0}v_{0}^{2}$ (thick line) and 
$p_{\rm ram}=10 \rho_{0}v_{0}^{2}$ (thin line).
The different inclination angles between the disk and the orbital
plane are: solid: 90$^{\rm o}$; dotted: 45$^{\rm o}$;
dashed: 20$^{\rm o}$; dash-dotted: 0$^{\rm o}$.
The vertical line at $t$=0~yr indicates the closest passage
of the galaxy to the cluster center. \label{fig:parameterplotting}}
\end{figure}

For the analysis of the simulation we concentrate on two time dependent
parameters which are of particular relevance for the evolution of the
galaxy:
\begin{itemize}
\item 
the total gas mass $M_{\rm tot}$ within the galaxy's radius $R$=20 kpc and 
a constant height $H$=1 kpc,
\item
and the central H{\sc i} surface density of the galaxy 
$\Sigma_{\rm central}$ with a spatial resolution of $\sim$2 kpc.
This is a slightly different definition than that of Cayatte et al. (1994)
(Table~\ref{tab:runs}), which loses its meaning if $D_{\rm HI} < D_{25}/4$.
\end{itemize}
The time dependence of these quantities is plotted in 
Fig.~\ref{fig:parameterplotting}. For clarity,  only the curves corresponding
to $p_{\rm ram}=10\ {\rm and}\ 100\rho_{0}v_{0}^{2}$ are shown.
The other curves are located between them.
The different inclinations angle $\Theta$ between the disk and the orbital
plane are plotted with different linestyles.
The maximum ram pressure strength of 100$\rho_{0}v_{0}^{2}$ corresponds to 
the lower thick graphs, the strength of 10$\rho_{0}v_{0}^{2}$ corresponds 
to the upper graphs in the same panel.
We recall here that in these simulations the maximum ram pressure
(i.e. the galaxy's closest approach to the cluster center) occurs at 
$t=0$~yr, indicated by the vertical line in Fig.~\ref{fig:parameterplotting}.

\subsection{The total gas mass \label{sec:totmass}}

Observationally this is the most important parameter because it
gives directly the H{\sc i} deficiency, which is defined as
\begin{equation}
DEF=<\log\,X>_{\rm T,D}-\log\,X_{\rm obs} 
\end{equation}
(Giovanelli \& Haynes 1985), where $X_{\rm obs}$ is the observed quantity
of the cluster galaxy and $<\log\,X>_{\rm T,D}$ is averaged over
a sample of field galaxies of morphological type $T$ and optical diameter $D$. 
$X$ can be the H{\sc i} mass, the H{\sc i} mass divided by the luminosity, or 
the H{\sc i} mass divided by the square of the optical linear diameter. In 
each case $X(T,D)$ is proportional to the H{\sc i} mass of a galaxy.
Since in our model the galaxy's diameter and luminosity do not change during
the simulations, we define the model H{\sc i} deficiency as follows:
\begin{equation}
DEF\ =\ {\rm log}(M_{\rm HI}^{\rm in}/M_{\rm HI}^{\rm f})\ ,
\label{eq:deficiency}
\end{equation}
where $M_{\rm HI}^{\rm in}$ and $M_{\rm HI}^{\rm f}$ are the total H{\sc i} 
masses at the beginning and the end of the simulation, i.e. after $t$=500~Myr.
We want to point out three important issues:

(i) In the model, ram pressure stripping acts very rapidly. 
For a cloud at rest,
the timescale to be accelerated to the escape velocity $v_{\rm esc}$ is 
given by Murakami \& Babul (1999):
\begin{equation}
\tau \sim \frac{v_{\rm esc} M_{\rm cl}}{p_{\rm ram} \pi R_{\rm cl}^{2}}\ ,
\end{equation}
where $M_{\rm cl}$ is the cloud mass, $R_{\rm cl}$ is the cloud radius,
and $P_{\rm ram}$ is the ram pressure of the hot ICM.
Assuming $v_{\rm esc}=\sqrt{2}\,v_{\rm Kepler} =$200 km\,s$^{-1}$,
$R_{\rm cl} =$10 pc, $M_{\rm cl} = 10^{4}$ M$_{\odot}$, and
$P_{\rm ram} = 100\,\rho_{0}\,v_{0}^{2}$ gives $\tau \sim 3\times 10^{7}$ yr.
This has also been pointed out by Abadi et al. (1999). 
However, for a Lorentzian temporal profile ram pressure acts over a long
period (up to several $10^{8}$~yr). Especially in the case of face-on stripping
($\Theta=90^{\rm o}$), where the restoring forces due to the galaxy's gravitational
potential are minimum, stripping begins $t \sim 300$~Myr before the closest
passage to the cluster center. For decreasing $\Theta$ this offset decreases.

(ii) The consequences of ram pressure stripping on the total gas H{\sc i} 
content can be observed earlier for higher maximum ram pressure values
and higher inclination angles (Fig.~\ref{fig:parameterplotting} and 
Table~\ref{tab:runs}). For a maximum ram pressure of 100$\rho_{0}v_{0}^{2}$ stripping 
begins $\sim 300$~Myr before ram pressure reaches its maximum, 
whereas for a maximum ram pressure of 20$\rho_{0}v_{0}^{2}$ and an inclination angle 
of $\Theta=0^{\rm o}$ stripping begins $\sim$50~Myr after ram pressure reaches its maximum.
The effects of ram pressure stripping on {\it non} H{\sc i} deficient 
galaxies can thus be observed in some cases before the galaxy's closest passage to the
cluster center.
In the case of nearly edge-on stripping $\Theta < 30^{\rm o}$ the gas in the outer
disk is mainly pushed to the inner part of the galactic disk in the
beginning of the stripping (this corresponds to the wings of the
Lorentzian profile). In this stage of the ISM--ICM interaction the gas
distribution in the galactic disk can be very asymmetric.

The earliest moment when an H{\sc i} deficient galaxy ($DEF \geq 0.3$) with a 
distorted gas distribution can be observed is $t \sim -170$~Myr (for face-on stripping).
If one assumes a constant mean velocity of $v_{\rm gal}=1700$~km\,s$^{-1}$
with respect to the ICM this corresponds to a minimum projected distance of 
$\sim 1^{\rm o}$ from the cluster center.
For $\Theta < 45^{\rm o}$ where there is a pronounced asymmetric gas distribution within
the galactic disk $t \sim -40$~Myr which translated to a minimum projected distance of 
$\sim 15'$.

\begin{figure}
	\resizebox{\hsize}{!}{\includegraphics{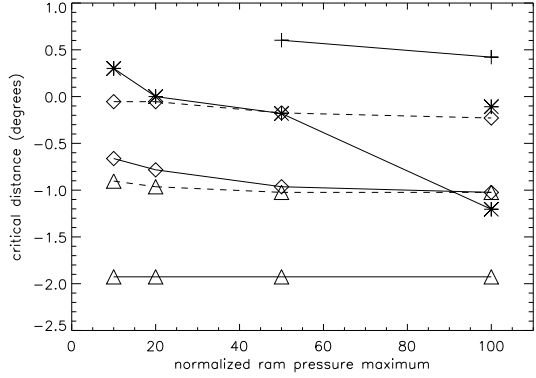}}
\figcaption[figure14.ps]{Critical distance at which
the total enclosed mass within 20~kpc drops below $5\times 10^{9}$~M$_{\odot}$
(solid lines) and $3\times 10^{9}$~M$_{\odot}$ (dashed lines) as a function 
of the normalized ram pressure maximum for each simulation.
Negative values indicate that the galaxy falls into the cluster center.
The different symbols correspond to the different inclination angles $\Theta$
(triangles: $\Theta =90^{\rm o}$, diamonds: $\Theta =45^{\rm o}$, stars: $\Theta =20^{\rm o}$,
crosses: $\Theta =0^{\rm o}$). \label{fig:striprad}}
\end{figure}

Fig.~\ref{fig:striprad} resumes the situation. The critical distance at which
the total enclosed mass within 20~kpc drops below $5\times 10^{9}$~M$_{\odot}$
(solid lines) and $3\times 10^{9}$~M$_{\odot}$ (dashed lines) is plotted
as a function of the normalized ram pressure maximum for each simulation.
Negative values indicate that the galaxy falls into the cluster center.
The different symbols correspond to the different inclination angles $\Theta$
(triangles: $\Theta=90^{\rm o}$, diamonds: $\Theta=45^{\rm o}$, stars: $\Theta=20^{\rm o}$,
crosses: $\Theta=0^{\rm o}$). The missing points or lines in the upper part of the
graph are due to the fact that the enclosed mass never drops below the
given limit, i.e. ram pressure is not strong enough. The maximum projected
distance at which a distortion of the gas distribution of a non H{\sc i} deficient 
galaxy can be observed when it falls into the cluster core is $\sim 2^{\rm o}$ 
(only in the case of face-on stripping).
The maximum distance for inclination angles $\Theta \leq 45^{\rm o}$ or H{\sc i} deficient 
galaxies is $\sim 1^{\rm o}$.

(iii) For low ram pressure maxima and low inclination angles a considerable 
amount of gas accretes back during 200--300~Myr
(Fig.~\ref{fig:parameterplotting}, Table~\ref{tab:runs}). This accretion sets in
at $t\sim$100--200~Myr after the closest approach to the
cluster center. In the case of edge-on stripping the accreted gas mass can attain 
values up to $5\times 10^{8}$~M$_{\odot}$. This corresponds to a maximum mass 
accretion rate of $\dot{M} \sim$1~M$_{\odot}$\,yr$^{-1}$. 
The fraction of the stripped gas which falls back to the galaxy depends
primarily on the inclination angle $\Theta$ between the disk and the orbital plane.
For one given $\Theta$ this fraction increases with decreasing maximum ram pressure.
In the case of edge-on stripping more than 50\% of the stripped gas can fall
back to the galaxy, whereas in the case of face-on stripping only
a few percent of the stripped gas re-accretes. The absolute value of the
gas which falls back onto the galaxy as a function of the maximum ram pressure
and the inclination angel $\Theta$ can be approximated crudely:
\begin{equation}
M_{\rm accr} \sim  6 \times \Big((\frac{\rho v^{2}}{\rho_{0} v_{0}^{2}})_{\rm max} \times 
\sin(\frac{9}{10}(\Theta+10^{\rm o}))^{2}\Big)^{-0.7}\ . 
\end{equation}
Thus, the re-accreted mass is approximately proportional to $(\rho v_{\perp}^{2})^{-0.7}$, 
where $v_{\perp}$ is the component of the galaxy's velocity within the cluster
perpendicular to the disk plane.

\subsection{\label{sec:censurfdens} The central surface density}

If one assumes a Schmidt law (Schmidt 1959, Kennicutt 1983) for the global star 
formation rate of a galaxy, the central surface density $\Sigma_{\rm c}$
is directly linked
to the star formation activity in the central part of the galactic disk. 
Due to the discreteness of the model and the
missing resolution in the galaxy center this parameter has
a large scatter. For all simulations $\Sigma_{\rm c}$
stays constant at the initial value until $t \sim -200$~Myr.
In the case of face-on stripping ($\Theta=90^{\rm o}$) 
$\Sigma_{\rm c}$ begins to decrease at $t \sim -100$~Myr.
It drops within $\Delta t \sim 150$~Myr to its minimum value and
then stays constant. The final central surface density decreases
with increasing maximum ram pressure. For the highest ram pressure
maximum ($p_{\rm ram}=100\,\rho_{0}v_{0}^{2}$)
it drops to about one third of its initial value. 

For $\Theta=45^{\rm o}$ $\Sigma_{\rm c}$ begins to rise
at $t \sim -100$~Myr, because a part of the gas is pushed
to smaller galactic radii in the beginning of the stripping.
The maximum which is about 1.2 times the initial value is reached
at $t \sim-100$ -- $-50$~Myr. For the lowest ram pressure
maximum ($p_{\rm ram}=10\,\rho_{0}v_{0}^{2}$) there is a second
increase and a second maximum (1.4 times the initial value)
at $t \sim 230$~Myr. For higher ram pressure maxima 
$\Sigma_{\rm c}$ decreases for $t > 0$~Myr.

For $\Theta=20^{\rm o}$ $\Sigma_{\rm c}$ begins to rise later
but more rapidly than for $\Theta=45^{\rm o}$ (at $t \sim -100$~Myr)
when the gas is pushed to smaller galactic radii.
The maximum is about 1.4 times the initial value. 
Only for the highest ram pressure maximum 
($p_{\rm ram}=100\,\rho_{0}v_{0}^{2}$) $\Sigma_{\rm c}$ drops
below its initial value for $t > 0$~Myr.
In the simulations with lower ram pressure maxima $\Sigma_{\rm c}$ 
has a secondary maximum at $t\sim 250$~Myr which has the same
value than the first one.

In the case of edge-on stripping ($\Theta=0^{\rm o}$) $\Sigma_{\rm c}$
begins to rise rapidly at $t \sim -50$~Myr and reaches its maximum
shortly after $t = 0$~Myr. The time when it reaches the maximum 
increases with decreasing ram pressure maximum.
The maximum of $\Sigma_{\rm c}$ increases with increasing
ram pressure maximum $p_{\rm ram}$. For $p_{\rm ram}=100\,\rho_{0}v_{0}^{2}$
it reaches 1.6 times its initial value.
As for $\Theta=20^{\rm o}$ there is a secondary maximum at $t \sim 320$~Myr
except for $p_{\rm ram}=100\,\rho_{0}v_{0}^{2}$. The offset
between the two maxima is constant $\Delta t \sim 300$~Myr,
i.e. the secondary maximum appears later for decreasing $p_{\rm ram}$.

In order to summarize our findings, we conclude that
the central surface density $\Sigma_{\rm c}$ decreases after the galaxy's
passage near the cluster center in the case of high
inclination angles (nearly face-on stripping). For
small inclination angles $\Theta$ (nearly edge-on stripping) $\Sigma_{\rm c}$
rises and reaches a maximum when the galaxy passes the cluster center.
In addition, there can be a secondary maximum after $\Delta t \sim 300$~Myr.

\section{Discussion}

\subsection{\label{subsec:comparison} Comparison with H{\sc i} observations}

\begin{figure}
	\resizebox{\hsize}{!}{\includegraphics{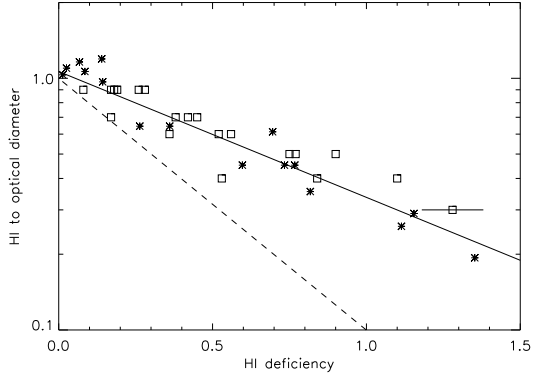}}
\figcaption[figure15.ps]{Normalized H{\sc i} to optical diameter as a function 
of the H{\sc i} deficiency. Squares: observed values 
(Cayatte et al. 1994); stars: model values. The horizontal
line represents the error in the H{\sc i} deficiency determination.
The solid line corresponds to $DEF \propto \log(D_{\rm HI}^{-2})$, the
dashed line corresponds to $DEF \propto \log(D_{\rm HI}^{-1})$.
\label{fig:cayatte94}}
\end{figure}

The results of our simulations can be directly compared with the H{\sc i} 
21 cm observations of Virgo cluster spirals made by Cayatte et al. (1994). 
We calculated the H{\sc i}
diameter in the same way as these authors: it is defined
by a limiting H{\sc i} column density of 10$^{20}$ cm$^{-2}$.
This parameter is plotted in Fig.~\ref{fig:cayatte94}. It
shows the normalized H{\sc i} to the optical diameter of our model together
with the observed values as a function of the H{\sc i} deficiency.
The model values are taken from the {\it final} states of our
simulations. We assumed an optical diameter of 20~kpc (which
corresponds to $\sim$7.5 disk scale lengths) for our model galaxy.
The initial H{\sc i} diameter is $D_{\rm HI}^{\rm init}$=30~kpc.

There is an excellent agreement between the model values and the
observational data for the normalized H{\sc i} to optical diameter.
We are able to reproduce the slope and the scatter of the observed data.

Now, if we assume that the H{\sc i} surface density  
has the same value before and after the stripping event, the final H{\sc i} mass is 
given by $M_{\rm HI}^{\rm fin} \simeq \pi \Sigma_{\rm HI} (R_{\rm HI}^{\rm fin})^{2}$,
where $R_{\rm HI}^{\rm fin}$ is the final H{\sc i} radius at the end
of a simulation. Thus the H{\sc i} deficiency depends on
the final H{\sc i} diameter in the following way:
$DEF \propto \log(D_{\rm HI}^{-2})$. This curve is represented 
by the solid line in Fig.~\ref{fig:cayatte94}. If we assume that
the H{\sc i} surface density is proportional to $R^{-1}$
and does not change during the stripping event, we obtain
$DEF \propto \log(D_{\rm HI}^{-1})$. This curve is represented 
by the dashed line in Fig.~\ref{fig:cayatte94}. Thus, the modeled
and observed H{\sc i} diameters as functions of the H{\sc i}
deficiency are consistent with a constant radial H{\sc i} surface density,
which has, in a first approximation, the same value before and after the stripping event.
However, Fig.~\ref{fig:parameterplotting} shows that the final
surface density can be decreased by up to a factor of $\sim$3 with respect to
the initial surface density.

Cayatte et al. (1994) divided the observed spiral galaxies into four
groups:

Group I: galaxies with $D_{\rm HI}/D_{0}\geq 1.3$, where $D_{\rm HI}$ and
$D_{0}$ are the H{\sc i} and optical diameters respectively.
These are cases of rather unperturbed H{\sc i} disks.

Group II: galaxies with 0.75$<D_{\rm HI}/D_{0}<1.3$. The surface density
falls off faster beyond the optical half radius.

Group III: galaxies with $D_{\rm HI}/D_{0}\leq 0.75$. The disks are strongly
truncated. 

Group IV: anemic galaxies with a central H{\sc i} hole.

\begin{figure*}[ht]
	\resizebox{\hsize}{!}{\includegraphics{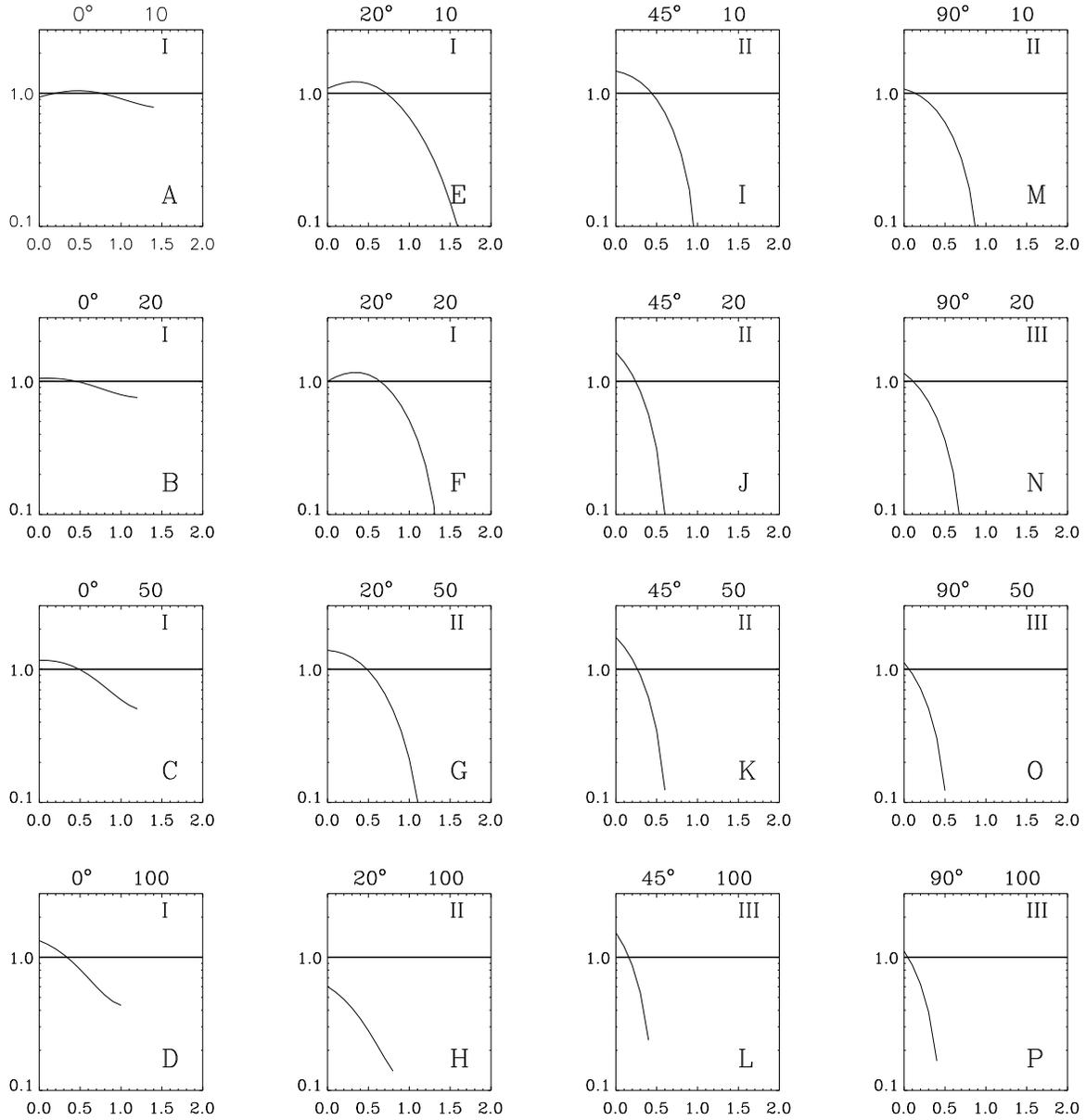}}
\figcaption[figure16.ps]{Smoothed normalized model surface density profiles.
The distance to the galaxy center is divided by the optical radius. 
The inclination angle between the galaxy's disk plane and its orbital 
plane together with the maximum ram pressure are indicated on top of 
each panel. The RUN is indicated in the lower right corner,
the group membership (Cayatte et al. 1994) in the upper right corner. 
 \label{fig:groups}}
\end{figure*}

Since the model is in principle not able to reproduce a central hole
which might be due to mechanisms other than ram pressure stripping,
we will not discuss group IV. Anemic galaxies will be discussed
in Section~\ref{sec:anemic}.
The model surface densities were fitted with a fourth order
polynomial to obtain smooth curves. This procedure can give
overestimated values of the central surface density which are
therefore not comparable to the value determined in 
Section~\ref{sec:censurfdens}. However, the overall behavior of the
profile (especially the cut off) is well reproduced.
We divided the final model surface density profiles by the initial one. 
These graphs are shown in Fig.~\ref{fig:groups}.
These normalized profiles can be directly compared with those of
Fig.~6 in Cayatte et al. (1994).
We can now assign each model density profile to one of the three groups.
This gives:\\
\\
Group I: RUN A(10,0), B(20,0), C(50,0), D(100,0), E(10,20), F(20,20)\\
Group II: RUN G(50,20), H(100,20), I(10,45), J(20,45), K(50,45), M(10,90)\\
Group III: RUN L(100,45), N(20,90), O(50,90), P(100,90)\\

The membership of a simulation to a group depends  
on the form of the orbit, i.e. on the maximum ram pressure and  
on the inclination angle between the galaxy's plane and the orbital plane. 
The comparison with the observed data shows that we can reproduce the 
scatter of the data over the whole range of deficiencies. The absolute
values of the surface density profiles
for the group I and III galaxies can be reproduced, whereas the values for the 
group II (intermediate deficiency) galaxies are higher than the observed values.
Only one simulation (H(100,20)) of a group II galaxy shows a significantly 
decreased final surface density ($\Sigma_{\rm HI} < 10^{21}$~cm$^{-2}$).
Thus, our data does not show a systematic decrease of the central H{\sc i} surface 
density after the stripping process for group II galaxy as observed by 
Cayatte et al. (1994). We suggest that this decrease
is related to the physical environment in the inner stellar and gaseous disk 
as the UV radiation field, the supernova rate, or the metalicity of the gas which 
influence the extent of the photodissociation regions and the gas heating and 
cooling rates leading to different H{\sc i} to H$_{2}$ ratios.

\subsection{Comparison with SPH simulations}

Abadi et al. (1999) made ram pressure stripping simulations using
an SPH code. They modeled a spiral galaxy moving with a constant
velocity within an ICM of constant density. This corresponds to a
circular galaxy orbit within the cluster. Therefore they do not
have re-accretion of material which is mainly due to the decrease of
ram pressure when the galaxy leaves the cluster core.
Their timescale for ram pressure stripping is between 1 and 5$\times 10^{7}$~yr,
thus comparable to our simulations. In difference to our model they have
used a decreasing gas surface density profile $\Sigma_{\rm gas} \propto R^{-1}$
that corresponds to the total gas surface density (atomic and molecular). 
They therefore underestimated the final H{\sc i} deficiency.
This becomes clear if one estimates the H{\sc i} deficiency using their
H{\sc i} to optical radius fraction and Fig.~\ref{fig:cayatte94}. For example,
$R_{\rm HI}/R_{\rm opt}\sim 0.3$ for their run D and E. According to
Fig.~\ref{fig:cayatte94} the resulting H{\sc i} deficiency is $DEF > 0.8$
compared to the given value of $DEF\sim 0.4$.

\subsection{Interpretation of the H{\sc i} data}

We conclude that our simulations can reproduce the observational data of 
Cayatte et al. (1994) in its behavior, absolute values, and scatter.
Therefore, we suggest that ram pressure stripping is the main cause of their 
H{\sc i} deficiency.

It is worth noting that we can reproduce the H{\sc i} observations using 
only one passage of the galaxy near the cluster center. 
As a typical orbital period of an H{\sc i} deficient
spiral galaxy is of the order of several Gyr, it is possible that
it has already passed through the cluster center more than once. 
The galaxy trajectory model has shown that the existence of
a perturbing gravitational potential (M86) in the cluster center has
two effects: (i) It can make the central mass
concentration (M87) move. As the intracluster gas will follow the
variable gravitational potential, the location of highest gas density
within the cluster can also change. (ii) The galaxy orbits will change
leading to a ram pressure profile changing with time for each passage
in the cluster center. 

Our simulations indicate (Fig.~\ref{fig:cayatte94}) that the lower H{\sc i}
surface densities of galaxies with $DEF\sim 0.5$ is due to internal
parameters of the galaxy (Section~\ref{subsec:comparison}) which lower
the H{\sc i} to H$_{2}$ ratio within the optical disk, whereas the total gas 
surface density stays the same. Furthermore, the simulations of Valluri (1993)
indicate that the gravitational potential of the galaxy does not
change significantly when the galaxy passes near the cluster center.
For the estimation of the final H{\sc i} deficiency after a stripping
event, the stellar and gas surface densities can thus be regarded as constant
in a first approximation. A galaxy with an already truncated H{\sc i} 
radius with an unchanged total gas surface density will only lose more gas 
in a further stripping event if the maximum ram pressure and/or
its inclination angle with respect to the orbital plane is higher than before.
If a galaxy has already accomplished several
orbits within the cluster, the orbit leading to the closest approach to 
the cluster center and/or having the highest inclination angle $\Theta$ is decisive 
for its H{\sc i} deficiency. On the other hand, if the final surface density
is decreased, a further passage through the cluster core will result
in a higher H{\sc i} deficiency.
The observed population of deficient galaxies represents galaxies which have
passed through the cluster center at least once. 

The different groups of Cayatte et al. (1994) represent galaxies 
on different orbits in the cluster. The fact that we can assign galaxies 
in the final state after the interaction to these groups confirms the 
conclusion that all H{\sc i} deficient galaxies have already passed through 
the cluster center.

The effects of ram pressure can be observed before the closest passage of
the galaxy to the cluster center only in the case of inclination angles
$\Theta \geq 45^{\rm o}$. The observed asymmetries of the gas distribution of 
H{\sc i} deficient ($DEF \geq 0.3$) in the Virgo cluster which 
are located more than 1$^{\rm o}$ from the cluster center
are due to a former ISM--ICM interaction. This interaction has already ceased but 
its consequences are still observable. This is also the case
for non H{\sc i} deficient galaxies which are located more than 2$^{\rm o}$
from the cluster center. Thus, these galaxies showing 
important distortions in their H{\sc i} distribution have already passed 
through the cluster center. They are {\it not} infalling galaxies.

\subsection{\label{sec:amount} The amount of stripping}

\begin{figure}
	\resizebox{\hsize}{!}{\includegraphics{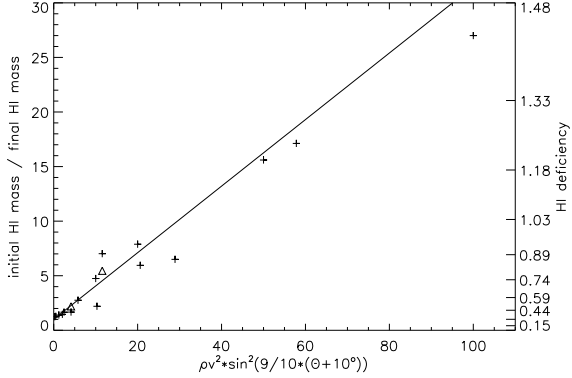}}
\figcaption[figure17.ps]{H{\sc i} deficiency as a function of the quantities 
$(\rho\,v^{2})/(\rho_{0}\,v_{0}^{2})$ (ram pressure strength) 
and $\Theta$ (inclination angle). The solid line represents a
linear least square fit. The triangles correspond to simulations
of galaxies with different initial H{\sc i} mass, initial H{\sc i}
diameter, and rotation curve.
\label{fig:strippingamount}}
\end{figure}

The knowledge of the dependence of the H{\sc i} deficiency 
on the maximum ram pressure
and the inclination angle together with the galaxy's radial velocity 
can constrain the galaxy orbits necessary to obtain a given H{\sc i} 
deficiency. We therefore attempt to give a simple analytical
formula. Fig.~\ref{fig:strippingamount} shows the fraction between the
final and initial total H{\sc i} mass
as a function of the quantities $(\rho\,v^{2})/(\rho_{0}\,v_{0}^{2})$
(ram pressure strength) and $\Theta$ (inclination angle).
The solid line represents a linear fit.  
The least square fit for the fraction between the initial and final total H{\sc i} mass is 
\begin{equation}
M_{\rm HI}^{\rm in}/M_{\rm HI}^{\rm f}=0.25\,\big(\frac{\rho\,v^{2}}{\rho_{0}\,v_{0}^{2}}\big)\,\sin^{2}(\frac{9}{10}(\Theta +10^{\rm o}))+0.84
\end{equation}

\subsection{\label{sec:mmassive} Galaxies with different properties}

The present simulations were made using a galaxy model with a constant
rotation curve of $v_{\rm rot}$=140~km\,s$^{-1}$, a total atomic gas
mass of $M_{\rm HI}^{\rm tot}=6 \times 10^{9}$~M$_{\odot}$, and an initial H{\sc i}
diameter of $D_{\rm HI}$=30~kpc. 
For a galaxy with different $v_{\rm rot}$, $M_{\rm HI}^{\rm tot}$, and
$D_{\rm HI}$ the resulting H{\sc i} deficiency changes in the
following way:

The final radius of the H{\sc i} gas distribution at the end of a simulation
is given by
\begin{equation}
R_{\rm HI}^{\rm fin} \simeq \Sigma_{\rm gas} p_{\rm ram}^{-1} v_{\rm rot}^{2}\ ,
\end{equation}
where $p_{\rm ram}=\rho_{\rm ICM}v_{\rm galaxy}^{2}$ and $\Sigma_{\rm gas}$
is the total gas surface density.
The fraction between the initial and final H{\sc i} mass is given by
\begin{equation}
\frac{M_{\rm HI}^{\rm init}}{M_{\rm HI}^{\rm fin}}=\frac{M_{\rm HI}^{\rm init}}
{\pi \Sigma_{\rm gas} (R_{\rm HI}^{\rm fin})^{2}}=\big(\frac{R_{\rm HI}^{\rm init}}
{R_{\rm HI}^{\rm fin}}\big)^{2}\ ,
\end{equation}
where $R_{\rm HI}^{\rm init/fin}$ is the initial/final H{\sc i} radius.
The total gas surface density can be regarded as constant (Section~\ref{subsec:comparison})
$\Sigma_{\rm gas}=\Sigma_{\rm gas}^{\rm init}$.
With $M_{\rm HI}=\pi \Sigma_{\rm gas}^{\rm init} (R_{\rm HI}^{\rm init})^{2}$
one obtains
\begin{equation}
\frac{M_{\rm HI}^{\rm init}}{M_{\rm HI}^{\rm fin}}=\frac{\pi^{2} p_{\rm ram}^{2}
(R_{\rm HI}^{\rm init})^{6}}{(M_{\rm HI}^{init})^{2} v_{\rm rot}^{4}}
\end{equation}
For a maximum ram pressure $p_{\rm ram}$, the fraction between the initial 
and final H{\sc i} mass for a galaxy with given $v_{\rm rot}$, $M_{\rm HI}^{\rm tot}$, 
and $D_{\rm HI}$ is
\begin{equation}
\frac{M_{\rm HI}^{\rm init}}{M_{\rm HI}^{\rm fin}}=\big(\frac{M_{\rm HI}^{\rm init}}
{M_{\rm HI}^{\rm fin}}\big)^{*} \times \big(\frac{R_{\rm HI}^{\rm init}}{30\ {\rm kpc}}
\big)^{6} \times \big(\frac{6\times 10^{9}\ {\rm M}_{\odot}}{M_{\rm HI}^{\rm init}}\big)^{2}
\times \big(\frac{140\ {\rm km}\,{\rm s}^{-1}}{v_{\rm rot}}\big)^{4}\ ,
\label{eq:fraction}
\end{equation}
where $\big(\frac{M_{\rm HI}^{\rm init}}{M_{\rm HI}^{\rm fin}}\big)^{*}$ is the
mass fraction for $v_{\rm rot}$=140~km\,s$^{-1}$, 
$M_{\rm HI}^{\rm tot}=6 \times 10^{9}$~M$_{\odot}$, and $D_{\rm HI}$=30~kpc.
In order to recover the deficiencies, one simply has to apply its definition
Eq.~\ref{eq:deficiency}.
We made two simulations with $v_{\rm rot}$=250~km\,s$^{-1}$, 
$M_{\rm HI}^{\rm tot}=8 \times 10^{9}$~M$_{\odot}$, and $D_{\rm HI}$=54~kpc:
(i) $\rho v^{2}=20 \rho_{0} v_{0}^{2}$ and $\Theta=20^{\rm o}$,
(ii) $\rho v^{2}=20 \rho_{0} v_{0}^{2}$ and $\Theta=45^{\rm o}$.
In order to verify Eq.~\ref{eq:fraction}, we calculated 
$\big(\frac{M_{\rm HI}^{\rm init}}{M_{\rm HI}^{\rm fin}}\big)^{*}$
using Eq.~\ref{eq:fraction} for these two simulations. The two resulting 
values can be seen as open triangles in Fig.~\ref{fig:strippingamount}. 
They are in good agreement with the mass fractions of the simulations 
of Table~\ref{tab:runs}.

\subsection{\label{sec:starform} Possible local star formation mechanisms}

The most important consequence of ram pressure stripping on cluster spiral 
galaxies is the removal of a large fraction of their gas reservoir.
Without the supply of fresh gas, the star formation activity of
a stripped galaxy will decrease on the timescale of a cluster crossing time.
Nevertheless, the interaction of the galaxy's gas with the ICM can
lead to a short enhancement of its star formation activity.
In this section we investigate possible star formation mechanisms
triggered by a ram pressure stripping event.
We first resume the relevant results of our simulations before discussing
their consequences on star formation activity.
\begin{itemize}
\item
A considerable fraction of the initial galaxy's gas mass can be accreted
back on the galaxy in the case of low ram pressure maxima and low 
inclination angles (up to 10\%).
\item
There is a strong enhancement of the central surface density 
for a high and intermediate maximum ram pressure and small
inclination angles $\Theta$.
\item 
In case of edge-on stripping and high maximum ram pressure, 
the central surface density increases by a factor 1.5 during the 
galaxy's passage in the cluster center.
\end{itemize}
If we relate the enhancement of the gas surface density to a local star formation rate
via the Schmidt law (Schmidt 1959, Kennicutt 1983) and via star formation due to
colliding flows (see e.g. Hunter et al. 1986),
these results lead to two possibilities of an enhancement of star formation:

(i) An instantaneous burst in the inner galactic disk which happens 
before ($\Delta t \leq 10^{8}$ yr) or while the galaxy passes through
the cluster center.
For the Virgo cluster this happens within a radius of $\sim 40'$
around the cluster center.
The enhancement of the star formation activity is strongest if the galaxy 
passes very close to the cluster center ($\sim$100 kpc) and if the
stripping is edge-on. Using a Schmidt-law, we estimate that the star formation 
rate within a radius $R= 0.25\times R_{25}$ can increase up to a factor $\sim$2.

(ii) The re-accretion of gas leads to star formation when the infalling
gas clouds collide with the clouds in the disk. This can lead to a
star formation event within the whole galactic disk which begins 
$\sim 2\times 10^{8}$ yr and ends up to
$\sim 5\times 10^{8}$ yr after the galaxy's closest passage to the cluster center.
In the case of maximum re-accretion (small $\Theta$, low maximum ram pressure)
$\sim 10\%$ of the total gas content falls back to the galaxy within 
$\sim3\times 10^{8}$~yr. If $\sim$10\% of this gas is turned into stars,
the resulting star formation rate is $SFR\sim 0.2$~M$_{\odot}$\,yr$^{-1}$.

We conclude that ram pressure stripping can induce two types of star formation
activity: 
a short burst during its passage in the cluster center or a longer lived
more continuous star formation activity due to the re-accretion of the
stripped gas.  The strength of the star formation activity is 
governed by the galactic orbits in the cluster. This confirms the suggestion of 
Bothun \& Dressler (1986) that ram pressure causes the H{\sc i} deficiency
and can induce star formation. Several individual
cases will be discussed in forthcoming papers.

\section{The influence of orbits on galaxy properties}

With the acquired knowledge of the previous section we can now
study the link between the properties of Virgo cluster spiral galaxies
and their orbits. Our model leads to the following scenario: 
galaxies on radial orbits will be H{\sc i} deficient regardless
of their morphology. These galaxies have already passed through the cluster
center at least once. For most of the group I and II galaxies, which have entered 
the cluster a few crossing times ago, the passage with the smallest impact parameter
was decisive for their H{\sc i} deficiency. 

Dressler (1986) and more recently Solanes et al. (2000) showed that the velocity 
dispersion of H{\sc i}
deficient cluster galaxies decreases with increasing distance to the cluster
center, indicating that these galaxies are on highly eccentric orbits.
In addition, they found indications for a segregation among the orbits of infalling 
spirals according to their H{\sc i} deficiency regardless of galaxy morphology. 
Interestingly, Solanes et al. (2000) pointed out that
the velocity dispersion of H{\sc i} deficient galaxies in the
Virgo cluster does not decrease with radius. They interpreted this result
as indicative for a gravitational field, which is strongly perturbed by
ongoing mergers of major subclumps. 

In order to investigate the link between H{\sc i} deficiency and galaxy
orbits, we use position--velocity plots of
different galaxy populations in the Virgo cluster together with our model orbits.

\subsection{H{\sc i} deficient galaxies}

\begin{figure}
	\resizebox{\hsize}{!}{\includegraphics{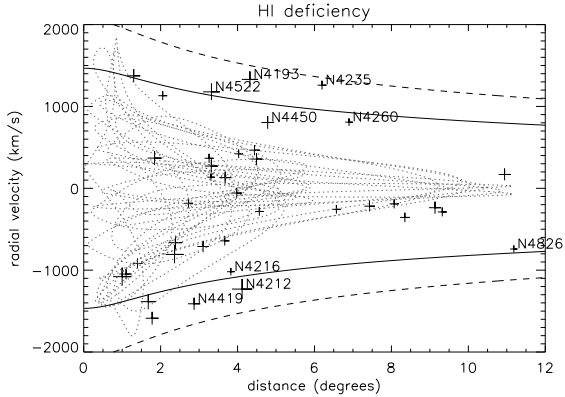}}
\figcaption[figure18.ps]{Radial velocity - projected distance plot for the 
H{\sc i} deficient ($DEF > 0.48$) Virgo cluster spirals. 
The curves of the model orbits (Section~\ref{sec:orbits}) are 
plotted as dotted lines, the maximum Keplerian velocity for the 
gravitational potential of M87 as solid lines. The size of the 
crosses is proportional to the H{\sc i} deficiency. \label{fig:deficiency}}
\end{figure}

We show the H{\sc i} deficiency data collected by Guiderdoni \&
Rocca-Volmerange (1986) in a radial velocity--projected distance plot
(Fig.~\ref{fig:deficiency}). 
The curves of the model orbits (Section~\ref{sec:orbits}) are plotted
as dotted lines, the Keplerian and escape velocity for the gravitational
potential of M87 as solid and dashed lines. 
The model orbits cover about half of the total area in the radial velocity--projected 
distance plane.
The size of the crosses is proportional
to the H{\sc i} deficiency. We only plot galaxies with $DEF\geq 0.48$.
9 out of 41 H{\sc i} deficient galaxies ($\sim20$\%) lie beyond the model orbits 
marked by the dotted lines for a maximum projected distance of 12$^{\rm o}$
(NGC~4193, NGC~4212, NGC~4216, NGC~4235, NGC~4260, NGC~4419, NGC~4450, NGC~4522, NGC~4826). 
The range of the projected distance of these galaxies is between 2$^{\rm o}$ and 
7$^{\rm o}$. Solanes et al. (2000) did not find a decrease of the velocity dispersion 
for distances smaller than 5$^{\rm o}$ because of these high velocity H{\sc i}
deficient galaxies. Their velocities are even higher than the Keplerian velocity
due to the M87 gravitational potential but still below the
escape velocity. Thus it is likely that they evolve on bound orbits. 
We will discuss this issue in Section~\ref{sec:infall}.
For larger radii we find a trend that H{\sc i} deficient galaxies
evolve on radial orbits as already pointed out by Dressler (1986) and Giraud (1986).

\subsection{\label{sec:anemic} Anemic galaxies}

\begin{figure}
	\resizebox{\hsize}{!}{\includegraphics{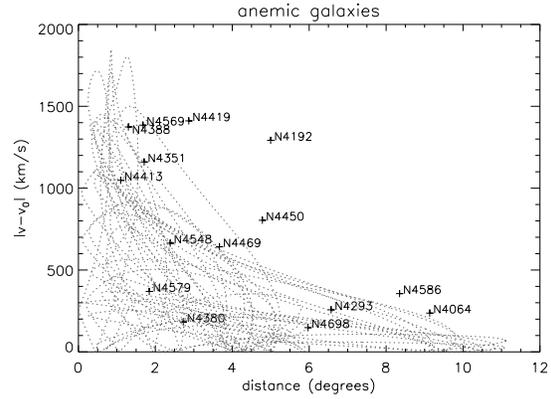}}
\figcaption[figure19.ps]{Radial velocity - projected distance plot for the 
anemic Virgo cluster spirals. 
The curves for for the model orbits (Section~\ref{sec:orbits}) are 
plotted as dotted lines. \label{fig:anemic}}
\end{figure}

Anemic galaxies are a special class of galaxies with a very low arm
inter-arm contrast (van den Bergh 1976). They are most similar
to the passive spirals observed in high redshift clusters
(see e.g.  Poggianti et al. 1999) and are believed to represent the oldest spiral 
galaxy population which has entered the cluster several Gyr ago. 
We suggest the following evolutionary scenario for these galaxies:
after a major stripping event the galaxy has lost its gas in the outer disk.
The ongoing star formation activity exhausted its remaining gas reservoir
in its entire disk. Without the possibility of a supply of fresh gas
from the outer disk, star formation activity in the spiral arm decreased
which led to a decreasing arm inter-arm contrast, i.e. the galaxy
became anemic. Thus, an anemic galaxy has entered the cluster
more than one crossing time ago on a radial orbit. In order to
investigate if anemic galaxies are still on radial orbits
or if tidal interactions have lead to more circular orbits, we show
the distribution of this population in the radial velocity--projected 
distance plot in Fig.~\ref{fig:anemic}. We include here also the group IV
galaxies of Cayatte et al. (1994), because we assume that the reduced
H{\sc i} diameter regardless of a central hole is due to the major stripping event.

The small number of orbits prevents us from a quantitative tracing
of the dispersion velocity as a function of the projected distance.
Nevertheless, we observe that 12 out of 15 anemic galaxies lie on radial orbits 
except (NGC~4192, NGC~4450, and NGC~4419). NGC~4192 
is seen nearly edge-on. Therefore, it is difficult to estimate its arm
inter-arm contrast. The low H{\sc i} deficiency of NGC~4192
indicates that this galaxy might be misclassified as an anemic galaxy.
If we exclude this galaxy, 
there is an indication that the velocity dispersion decreases with increasing 
distance. This is consistent with anemic galaxies being preferentially on radial 
orbits. As they represent a spiral population which entered the cluster
more than several Gyr ago, there is an indication that their orbits
were not perturbed by the changing gravitational field due to the merging of subclumps.
If this tendency turns out to be real, this would mean that only the infalling
galaxies are scattered with high velocities to large distances from the cluster center
during the merging of a major subclump (e.g. M86) and not the old spiral population.
Interestingly, Solanes et al. (2000) showed that the velocity dispersion of early type
spiral galaxies does not decrease. Anemic galaxies might thus form a special class
of galaxies whose orbits are of different shape than those of early type spirals.

\section{\label{sec:infall} Infall}

We will now try to explain the existence of the group of high velocity 
H{\sc i} deficient galaxies which can be found at projected distances of 
$\sim4^{\rm o}$ (NGC~4193, NGC~4212, NGC~4216, NGC~4235, NGC~4260, NGC~4419, 
NGC~4450, NGC~4522). 
One of its members (NGC 4522) has been observed in H$\alpha$
by Kenney \& Koopmann (1999). They claim that there are extraplanar ionized
filaments and identify it as one of the best candidates for ram pressure
stripping. As we have shown before, spiral galaxies on radial orbits
which lead to a strong stripping event do not permit these positions
in the radial velocity - projected distance plot.
Thus, there must be another explanation.

The basic idea comes from the presence of M86 in the vicinity of M87.
Since it has a high radial velocity with respect to M87, it is 
probable that M86 is falling into the cluster from behind. 
In addition, the central dwarf population shows already a spatial distortion
in the direction of M86 (Schindler et al. 1999). We interpret this as the 
consequence of the passage of M86 near the cluster center, i.e. M87.
If the subclump of M86 has entered the Virgo cluster with its own
spiral galaxy population, these could have been captured by the Virgo cluster.
The outcoming orbits of this infalling spiral population can
be very different from those which are already settled.

We have tested this hypothesis with the help of our orbit simulations.
M87 has been placed at rest in the center. M86 is located at (0.5 Mpc, 0 Mpc,
5 Mpc) and with an initial velocity vector
(0 km\,s$^{-1}$, 1 km\,s$^{-1}$, 0 km\,s$^{-1}$). For the test
galaxy we adopted: ${\bf x}$ = (0.5 Mpc, 1 Mpc, 5 Mpc) and ${\bf v}=$
(258 km\,s$^{-1}$, 1 km\,s$^{-1}$,  258 km\,s$^{-1}$), i.e. the test galaxy 
is orbiting at a distance of 1 Mpc around M86 which is
falling radially (along the LOS) into the gravitational potential of M87.
We have included the full symmetric gravitational potential of both galaxies.
The effects of an internal perturbation of M86 due to the tidal interaction with 
M87 is neglected. The assumption is justified, because the difference between 
their radial velocities $\Delta v \simeq 1500$~km\,s$^{-1}$ is much larger 
than the dispersion velocity of M86.

\begin{figure}
	\resizebox{5cm}{!}{\includegraphics{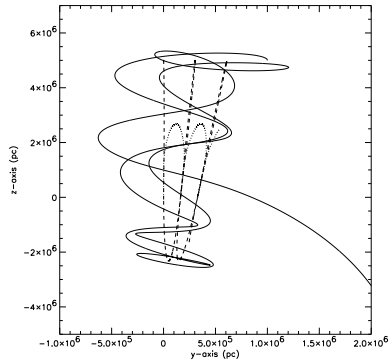}}
\figcaption[figure20.ps]{Trajectories of the test galaxy orbiting around M86
which is falling radially into the gravitational potential
of M87. Solid line: test galaxy. Dotted line: M87. Dashed line: M86.
\label{fig:infall_yz}}
\end{figure}

\begin{figure}
	\resizebox{\hsize}{!}{\includegraphics{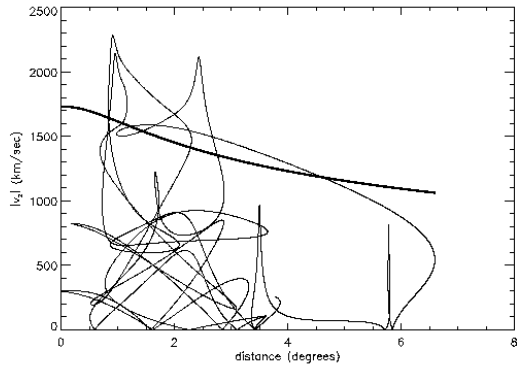}}
\figcaption[figure21.ps]{Radial velocity - projected distance plot for the
trajectories of Fig.~\ref{fig:infall_yz}. The Keplerian
velocity due to the gravitational potential of M87
is plotted as a thick line. The positions and velocities are 
plotted with respect to the positions and velocities of M87.
 \label{fig:infall_velocity}}
\end{figure}

The resulting trajectories can be seen in Fig.~\ref{fig:infall_yz}.
When the M86 clump approaches M87 the latter also begins to move.
The test galaxy is orbiting in the joint gravitational potential of M87
and M86 following M86 in the $z$-direction. When passing near M87
it can even be ejected from the cluster as it happens at the end of our
simulation. The corresponding radial velocity - projected distance plot
can be seen in Fig.~\ref{fig:infall_velocity}. The positions and
velocities are plotted with respect to the positions and velocities
of M87, the center of the Virgo cluster.
Clearly there is the possibility to obtain high radial velocities
at large projected distances especially when the galaxy is ejected 
from the cluster. There is also the effect that M87 begins to move,
changing the position of the cluster center. 

We suggest that
the existence of the special galaxy population with high radial 
velocities at large projected distances is linked to the trajectory of M86
perturbing the gravitational potential. In this scenario these galaxies
pass the cluster center with high velocities. 
The increased velocity leads to an enhanced ram pressure 
($p_{\rm ram} \propto v_{\rm gal}^{2}$). 
Solanes et al. (2000) also suggested
that gravitational perturbations due to major subclumps (as M86) are 
responsible for the high velocities H{\sc i} deficient galaxies.
If these galaxies are bound, they will evolve on extremely eccentric
orbits with very long periods. After the merging of the subclump, they will
be most probably found at very large distances from the cluster center and
the velocity dispersion of the remaining H{\sc i} deficient galaxies will again 
decrease.

Since there are indications that galaxies are accreted in subclumps, the
accretion rate is expected to be highly irregular showing strong peaks 
when a subclump as M86 is falling into the cluster adding new spiral galaxies
to the existing population.

\section{Galaxy evolution within the cluster}

Accreted spiral galaxies are preferentially
on radial orbits which lead them near the cluster center
where ram pressure is most efficient. 
Very recently accreted galaxies ($\sim$1 Gyr) can have an enhanced
star formation rate in the outer disk
due to the re-accretion of stripped gas. During the ram pressure stripping
event, the atomic gas is stripped from the outer disk leaving the
molecular gas unaffected. Consequently the star formation activity in the
outer disk region of highly stripped galaxies decreases, 
whereas the star formation rate in the
inner disk stays roughly constant after a possible short increase 
(see Section~\ref{sec:starform}). This leads to a decrease of the overall star 
formation rate. Koopmann \& Kenney (1998) noticed that the morphological
classification of the spiral galaxies in the Virgo cluster taking only
into account the overall star formation rate is often ambiguous.
Galaxies classified in this way as early type spirals become late types
if one takes into account their disk to bulge ratios in the R band. These galaxies
show a reduced overall star formation rate, which is mainly due to a strongly 
reduced star formation rate in the outer disk. We suggest that this special
type of galaxies has undergone a ram pressure stripping event.

We will now estimate the influence of ram pressure stripping on the global
star formation rate.
We assume a gas surface density distribution 
$\Sigma_{\rm H_{2}}=\Sigma_{0}\,\exp(-R/a)$ for the molecular phase and 
$\Sigma_{\rm HI}={\rm const}$ for the atomic phase within an optical radius.
Furthermore, the star formation rate is given by a Schmidt law (Kennicutt 1998)
\begin{equation}
SFR=2\,\pi\,2.5\times 10^{-10} \int_{0}^{R_{1}} R\,\Sigma_{\rm tot}^{1.4}\,dR\ ,
\end{equation}
where $\Sigma_{\rm tot}=\Sigma_{\rm H_{2}}+\Sigma_{\rm HI}$. 
It is assumed that star formation takes only place within the optical radius. 
After a major stripping event, the H{\sc i} is $R_{\rm HI}=0.3\times R{\rm opt}$.
We consider two different cases:

(i) $\Sigma_{0}=100$~M$_{\odot}$pc$^{-2}$, $a=2$~kpc, 
$\Sigma_{\rm HI}$=5~M$_{\odot}$pc$^{-2}$. For $R_{1}=R_{25}$ the total molecular gas mass
and the total atomic gas mass within the optical radius are comparable
$M_{\rm H_{2}}\simeq M_{\rm HI}$. The star formation drops by a factor 2
due to ram pressure stripping (from 3.6 to 1.7~M$_{\odot}$\,yr$^{-1}$).

(ii) $\Sigma_{0}=50$~M$_{\odot}$pc$^{-2}$, $a=2$~kpc, 
$\Sigma_{\rm HI}$=5~M$_{\odot}$pc$^{-2}$. In this case 
$M_{\rm H_{2}}\simeq 0.5\times M_{\rm HI}$. The star formation drops almost by a 
factor 3 due to ram pressure stripping (from 2.2 to 0.8~M$_{\odot}$\,yr$^{-1}$).

For this high maximum ram pressure star formation induced by re-accretion is negligible. 
In the case of a central starburst triggered by ram pressure, we assume an
increase of the local star formation rate within $R<0.25\,R_{25}$ by a factor 4 during
$5\times 10^{7}$~yr (Fig.~\ref{fig:parameterplotting}. 
This leads to a mild increase of the global star formation rate
(up to a factor 2)). The consumed molecular gas masses for both cases
are (i) $M_{\rm cons}\sim 3\times 10^{8}$~M$_{\odot}$ and (ii) 
$M_{\rm cons}\sim 1.3\times 10^{8}$~M$_{\odot}$. This corresponds to roughly
10\% of the total molecular gas content in both cases. Thus, the occurrence of a central
star burst can additionally lower the global star formation rate by $\sim$10\% 
after a strong stripping event.

If the gas content within an optical radius of a non deficient galaxy 
is dominated by the atomic phase, ram pressure stripping can
account for the quenching of star formation at a timescale of $\sim$1 Gyr described 
by Poggianti et al. (1999). In the case of a strong edge-on stripping event,
the global star formation will increase by a factor $\sim$1.5 within $\sim 10^{7}$~yr
and drop then by a factor 2--3 three with respect to its initial value. 
On the contrary, ram pressure cannot account for the morphological transformation,
which has a larger timescale (several Gyr). This transformation
might be due to the combined effect of high speed encounters with
massive galaxies and the interaction with the gravitational potential
of the cluster (``galaxy harassment''; Moore et al. 1996, 1998).

We will now attempt to determine the galaxies' location in the $z$
direction with respect to the cluster center (M87) using the following
results from this article:
\begin{itemize}
\item
If an H{\sc i} deficient galaxy shows enhanced star formation in the 
disk due to re-accretion of the stripped material, 
it has been accreted recently (less than 1 Gyr) by the cluster. We 
observe the galaxy when it has already passed the cluster center. It is in 
the phase of re-accretion of the stripped material which collides with
the galaxy's ISM leading to an enhancement of its star formation activity.
\item
If an H{\sc i} deficient galaxy shows a perturbed H{\sc i} velocity field
or a distorted H{\sc i} distribution, it has passed the
cluster center recently and is now coming out of the cluster center.
\item 
If a non H{\sc i} deficient galaxy shows a perturbed H{\sc i} velocity field
or a distorted H{\sc i} distribution and is located more than 2$^{\rm o}$
away from the cluster center, it has passed the
cluster center recently and is now coming out of the cluster center.
\end{itemize}
These galaxies have in common that they are 
coming from the cluster core with the closest approach to the cluster
center less than 1 Gyr ago. The sign of their radial velocity 
gives us the clue to place them with respect to M87. If the galaxy's
radial velocity is positive it is located behind the cluster
center, if it is negative it is located in front.

\section{Summary and conclusions}

In this work we have made N-body simulations in order to simulate 
the neutral gas content of spiral galaxies entering a cluster. 
We are interested in the effect 
of the ram pressure exerted by the hot intracluster medium on the ISM
of the fast moving galaxy. At the same time we simulated the gas dynamics
of a spiral galaxy on a radial orbits within the gravitational potential 
of the Virgo cluster. We investigated different orbits varying systematically
the inclination angle $\Theta$ between the disk and the orbital plane.
The main results are:
\begin{enumerate}
\item
The amount of the stripping and thus the H{\sc i} deficiency depends on the
galaxy orbit (minimum distance to the cluster center, maximum velocity, and 
inclination of the disk with respect to the orbital plane). We generalize
the formula of Gunn \& Gott (1972) including explicitly the inclination angle.
We give an approximation
of the relation between the H{\sc i} deficiency on the one hand and the maximum 
ram pressure and inclination angle on the other hand.
\item
We have used a realistic temporal ram pressure profile, i.e. ram pressure
decreases after the closest passage to the cluster center.
For maximum ram pressure of $p_{\rm ram} \simeq 1000-5000$~cm$^{-3}$\,(km\,s$^{-1}$)$^{2}$
and low inclination angles $\Theta <45^{\rm o}$ we observe a phase of re-accretion
where the gas, which has not been accelerated to the escape velocity, falls
back to the galaxy. The fraction of the re-accreted mass can attain up
to 10\% of the total atomic gas mass.
\item
For maximum ram pressure of $p_{\rm ram}\geq 2000$~cm$^{-3}$\,(km\,s$^{-1}$)$^{2}$
and small inclination angles $\Theta < 20^{\rm o}$ the central gas surface density can
increase up to a factor 1.5 within a few 10$^{7}$yr. Assuming a Schmidt
law for the star formation rate leads to an increase of the local star formation
rate up to a factor $\sim$2.
\item
Deformations of the neutral gas content are mainly observable {\it after}
the closest approach to the Virgo cluster center. Thus, the majority of
galaxies with a peculiar H{\sc i} distribution are coming out of 
the cluster core. 
\item
The observationally established correlation between the H{\sc i} to optical diameter
and the H{\sc i} deficiency is well reproduced by our simulations.
\end{enumerate}
We conclude that the scenario where the H{\sc i} deficiency is due to ram pressure 
stripping is consistent with all available H{\sc i} 21~cm observations of the Virgo 
spiral galaxy population.

\begin{acknowledgements}
We would like to thank P. Maloney for providing the calculations in 
Section~\ref{sec:fate} and the anonymous referee for helping us
to improve this paper considerably. 
BV was supported by a TMR Programme of the European Community
(Marie Curie Research Training Grant). 
\end{acknowledgements}

\clearpage

\clearpage

\clearpage

\clearpage

\clearpage

\clearpage

\clearpage

\begin{deluxetable}{lccccccccc}
\tablecaption{Parameters and results of the different simulations 
\label{tab:runs}}
\tablehead{
\colhead{RUN} & 
\colhead{$\big(\frac{\rho\,v^{2}}{\rho_{0}\,v_{0}^{2}}\big)_{\rm max}^{\rm a}$} &
\colhead{$\Theta ^{\rm b}$} &
\colhead{$\Delta t^{\rm c}$} &
\colhead{$\Delta D^{\rm d}$} &
\colhead{$D_{\rm HI}^{\rm e}$} &
\colhead{$\Sigma_{\rm central}^{\rm f}$} &
\colhead{DEF$^{\rm g}$} &
\colhead{M$_{\rm accr}^{\rm h}$} &
\colhead{M$_{\rm accr}/M_{\rm str}\ ^{\rm i}$}
}
\startdata
A(10,0)&10&0&-&-&32&12.71&0.01&2.4&1.41\\
B(20,0)&20&0&-&-&34&13.85&0.02&3.1&0.94\\
C(50,0)&50&0&100&180&36&15.31&0.05&5.7&0.68\\
D(100,0)&100&0&70&125&37&12.76&0.10&5.1&0.32\\
\hline
E(10,20)&10&20&50&90&33&15.87&0.08&1.8&0.17\\	
F(20,20)&20&20&0&0&30&16.26&0.13&2.5&0.15\\
G(50,20)&50&20&-30&-50&20&16.57&0.23&1.0&0.04\\
H(100,20)&100&20&-200&-360&19&4.9&0.61&0.4&0.01\\
\hline
I(10,45)&10&45&-110&-200&20&17.68&0.34&2.6&0.08\\
J(20,45)&20&45&-130&-230&14&18.03&0.73&1.5&0.03\\
K(50,45)&50&45&-160&-285&14&10.93&0.72&0.8&0.02\\
L(100,45)&100&45&-170&-300&9&4.83&1.13&0.3&0.01\\
\hline
M(10,90)&10&90&-320&-570&14&13.12&0.60&2.3&0.05\\
N(20,90)&20&90&-320&-570&11&10.75&0.82&1.2&0.02\\
O(50,90)&50&90&-320&-570&8&5.64&1.11&0.5&0.01\\
P(100,90)&100&90&-320&-570&6&3.33&1.35&0.3&0.01\\
\enddata
\tablenotetext{a}{ Maximum ram pressure in units of $\rho_{0}v_{0}^{2}$.} 
\tablenotetext{b}{ Inclination angle with respect to the orbital plane in 
degrees.}
\tablenotetext{c}{ Time between the beginning of the stripping (total gas 
mass within 20~kpc $M_{\rm tot} < 5\times10^{9}$~M$_{\odot}$) and the
galaxy's closest passage to the cluster center in Myr.}
\tablenotetext{d}{ Radial distance in kpc corresponding to $\Delta t$ assuming 
a constant velocity of the galaxy with respect to the cluster center of 
$v_{\rm gal}$=1700 km\,s$^{-1}$.} 
\tablenotetext{e}{ Final H{\sc i} diameter in kpc.}
\tablenotetext{f}{ Final central surface density in M$_{\odot}$\,pc$^{-2}$.}
\tablenotetext{g}{ Final H{\sc i} deficiency.}
\tablenotetext{h}{ Re-accreted gas mass in $10^{8}$~M$_{\odot}$.}
\tablenotetext{i}{ Re-accreted gas mass divided by the stripped gas mass.}
\end{deluxetable}


\begin{thebibliography}{}

\bibitem{q1} Abadi, M.G., Moore, B., \& Bower, R.G. 1999, \mnras, 308, 947
\bibitem{q2} Allen, C., \& Santill\'an, A. 1991, RMAA, 22, 255
\bibitem{q4} Balsara, D., Livio, M., \& O'Dea, C.P. 1994, \apj, 437, 83
\bibitem{q5} Barnes, J.E., \& Hut, P. 1986, \nat, 324, 446
\bibitem{q6} Bicay, M.D., \& Giovanelli, R. 1987, \apj, 321, 645
\bibitem{q7} Binney, J., \& Tremaine, S. 1987, Galactic Dynamics (Princeton: Princeton University Press), p. 77
\bibitem{q8} B\"{o}hringer, H., Briel, U.G., Schwarz, R.A., Voges, W., Hartner, G., \& Tr\"{u}mper, J., 1994, \nat, 368, 828
\bibitem{q9} Boselli, A., Gavazzi, G., Lequeux, J., Buat, V., Casoli, F., Dickey, J., \& Donas, J. 1997, \aap, 327, 522
\bibitem{q10} Bothun, G., \& Dressler A., 1986, \apj, 301, 57
\bibitem{q11} Bothun, G., Schommer, R.A., \& Sullivan W.T.III 1982, \aj, 87, 731
\bibitem{q12} Byrd, G., Valtonen, M. 1990, \apj, 350, 89
\bibitem{q13} Caldwell, N., Rose, J.A., Sharples, R.M., Ellis, R.S., \& Bower, R.G. 1993, \aj, 106, 473
\bibitem{q14} Caldwell, N., \& Rose, J.A. 1997, \aj, 113, 492
\bibitem{q15} Caldwell, N., Rose, J.A., \& Dendy, K., 1999, \apj, 117, 140
\bibitem{q16} Cavaliere, A., \& Fusco-Femiano, R., 1976, \aap, 49, 137
\bibitem{q17} Cayatte, V., van Gorkom, J.H., Balkowski, C., \& Kotanyi C. 1990, \aj, 100, 604
\bibitem{q18} Cayatte, V., Kotanyi, C., Balkowski, C., \& van Gorkom, J.H. 1994, \aj, 107, 1003
\bibitem{q19} Chamaraux, P., Balkowski, C., \& G\'erard, E. 1980, \aap, 83, 38
\bibitem{q20} Combes, F., \& G\'erin, M. 1985, \aap, 150, 32
\bibitem{q21} Couch, W.J., Barger, A.J., Smail, I., Ellis, R.S., \& Sharples, R.M. 1998, \apj, 497, 188 
\bibitem{q22} Cowie, L.L., \& McKee, C.F., 1977, \apj, 211, 135
\bibitem{q23} Cowie, L.L., McKee, C.F., \& Ostriker J.P., 1981, \apj, 247, 908
\bibitem{q24} Dressler, A. 1986, \apj, 301, 35
\bibitem{q25} Dressler, A., Smail, I., Poggianti, B.M., Butcher, H., Couch, W.J., Ellis, R.S., \& Oemler A.Jr. 1999, \apjs, 122, 51
\bibitem{q26} Gaetz, T.J., Salpeter, E.E., \& Shaviv G. 1987, \apj, 316, 530
\bibitem{q27} Gavazzi, G., Boselli, A., \& Kennicutt, R. 1991, \aj, 101, 1207
\bibitem{q28} Gavazzi, G., Catinella, B., Carrasco, L., Boselli, A., \& Contursi, A. 1998, \aj, 115, 1745
\bibitem{q30} Ghigna, S., Moore, B., Governato, F., Lake, G., Quinn, T., \& Stadel, J. 1998, \mnras, 300, 146
\bibitem{q31} Giovanelli, R., \& Haynes, M.P. 1985, \apj, 292, 404
\bibitem{q32} Giraud, E. 1986, \aap, 167, 25
\bibitem{q33} Guiderdoni, B., \& Rocca-Volmerange, B. 1985, \aap, 151, 108
\bibitem{q34} Gunn, J.E., \& Gott, J.R. 1972, \apj, 176, 1
\bibitem{q35} Hoffman, G.L., Helou, G., Salpeter, E.E., 1988, \apj, 324, 75
\bibitem{q36} Hollenbach, D.J., Tielens A.G.G.M., 1997, AARA, 35, 179
\bibitem{q37} Huchra, J.P. 1988, in The extragalactic distance scale, ed. S. van den Bergh \& C.J. Prichet, ASP Conference Series, Vol. 4 (Salt Lake City: ASP), p. 257
\bibitem{q38} Hunter, J.H.Jr. ,Sandford, M.T.II, Whitaker, R.W., Klein, R.I., 1986, \apj, 305, 309
\bibitem{q39} Kenney, J.D., \& Koopmann, R.A. 1999, \aj, 117, 181
\bibitem{q40} Kenney, J.D., \& Young, J.S. 1988, \apjs, 66, 261
\bibitem{q41} Kenney, J.D., \& Young, J.S. 1989, \apj, 344, 171
\bibitem{q42} Kennicutt, R.C.Jr. 1983, \aj, 88, 483
\bibitem{q43} Kennicutt, R.C.Jr. 1998, ARAA, 36, 189
\bibitem{q44} Kim, S., Staveley-Smith, L., Dopita, M.A., Freeman, K.C., Sault, R.J., Kesteven, M.J., \& McConnell D. 1998, \apj,  503, 674
\bibitem{q45} Klari\'c, M. 1995, \aj, 109, 2522
\bibitem{q46} Knude, J. 1981, \aap, 98, 74
\bibitem{q47} Koopmann, R.A., \& Kenney, J.D. 1998, \apj, 497, 75
\bibitem{q48} Kulkarni, S.R., \& Heiles, C., 1988, in Galactic and Extragalactic Radio
Astronomy, ed. G.L. Verschuur \& K.I. Kellermann (New York: Springer Verlag), 95
\bibitem{q49} Maloney, P.R., Hollenbach, D.J., \& Tielens, A.G.G.M., 1996, \apj, 466, 561 
\bibitem{q50} Moore, B., Katz, N., Lake, G., Dressler, A., \& Oemler A. 1996, \nat, 379, 613
\bibitem{q51} Moore, B., Lake, G., \& Katz, N. 1998, \apj, 495, 139 
\bibitem{q52} Murakami, I., \& Babul, A. 1999, \mnras, 309, 161
\bibitem{q53} Phookun, B., \& Mundy, L.G., 1995, \apj, 453, 154
\bibitem{q54} Poggianti, B.M., Smail, I., Dressler, A., Couch, W.J., Barger, A.J., Butcher, H., Ellis, R.S., \& Oemler A.Jr. 1999, \apj, 518, 576
\bibitem{q55} Rots, A.H., Crane, P.C., Bosma, A., Athanassoula, E., \& van der Hulst J.M. 1990, \aj, 100, 387
\bibitem{q56} Sanders, D.B., Scoville, N.Z., \& Solomon, P.M. 1985, \apj, 289, 373
\bibitem{q57} Schindler, S., Binggeli, B., \& B\"{o}hringer, H. 1999, \aap, 343, 420
\bibitem{q58} Schmidt, M., 1959, \apj, 129, 243
\bibitem{q59} Solanes, J.M., Manrique, A.,  Garc\'{\i}a-G\'omez, C., Gonz\'alez-Casado, G., Giovanelli, R., \& Haynes, M., 2000, \apj, in press
\bibitem{q59a} Spitzer L., 1978, Physical Processes in the Interstellar Medium
\bibitem{q60} Stoer, J., \& Burlisch, R. 1980, Introduction to Numerical Analysis (New York: Springer-Verlag), chapter 7
\bibitem{q60a} Takeda, H., Nulsen, P.E.J., \& Fabian, A.C. 1984, \mnras, 208, 261 
\bibitem{q61} Tosa, M. 1994, \apj, 426, L81 
\bibitem{q62} Tully, R.B., \& Shaya, E.J. 1984, \apj, 281, 31
\bibitem{q64} Valluri, M. 1993 \apj, 408, 57
\bibitem{q65} van den Bergh, S. 1976, \apj, 206, 883
\bibitem{q66} van Gorkom, J.H., \& Kotanyi, C.G. 1985, in Proceedings of the Workshop on the Virgo Cluster, edited by O.G. Richter and B. Bingelli (ESO Garching), p. 61
\bibitem{q66a} Vollmer B., Marcelin M., Amram P., Balkowski C., Cayatte V., Garrido O. 2000, \aap, 364, 532
\bibitem{q66b} Vollmer B., Braine J., Balkowski C., Cayatte V., Duschl W.J. 2001, \aap, accepted for publication
\bibitem{q67} Warmels, R.H. 1988, A\&AS, 72, 19
\bibitem{q68} Wiegel, W. 1994, Diploma thesis, University of Heidelberg
\bibitem{q70} Zabludoff, A.I., Zaritsky, D., Lin, H., Tucker, D., Hashimoto, Y., Shectman, S.A., Oemler, A., Kirshner, R.P., 1996, \apj, 466, 104
\end{thebibliography}
\end{document}